\documentclass[showpacs,amsmath,twocolumn,showkeys,pra]{revtex4-1}
\usepackage{dcolumn}    
\usepackage{bm}         
\usepackage{ifpdf}
\usepackage{amssymb,lineno,amsfonts}
\usepackage{graphicx}   
\usepackage{bbm}
\usepackage{mathrsfs}
\usepackage{upgreek}
\usepackage{epstopdf}
\usepackage{setspace}
\usepackage{hyperref}
\usepackage{float}
\usepackage{natbib}
\usepackage{color} 
\usepackage{subfigure}
\usepackage{comment}
\definecolor{med-blue}{RGB}{25,25,112} 
\hypersetup{colorlinks, linkcolor={blue},citecolor={blue}, urlcolor={blue}}
\newcommand{\be}{\begin{equation}}
\newcommand{\ee}{\end{equation}}
\newcommand{\bea}{\begin{eqnarray}}
\newcommand{\eea}{\end{eqnarray}}
\usepackage{tabularx}

\begin{document}
\title{Photon statistics of a double quantum dot micromaser: Quantum treatment}


\author{Bijay Kumar Agarwalla}
\affiliation{Department of Physics, Dr. Homi Bhabha Road, Indian Institute of Science Education and Research, Pune, 411008 India}
\author{Manas Kulkarni}
\affiliation{International Center for theoretical Sciences, Tata Institute of Fundamental Research, Bangalore-560089, India}

\author{Dvira Segal}
\affiliation{Chemical Physics Theory Group, Department of Chemistry,
and Centre for Quantum Information and Quantum Control,
University of Toronto, 80 Saint George St., Toronto, Ontario, Canada M5S 3H6}

\date{\today}
\begin{abstract}
A semiconductor single-atom micromaser consists of a microwave cavity coupled to a gain medium,
a double quantum dot driven out of equilibrium by a bias voltage.
The masing threshold of this system was recently probed
by measuring photon statistics in the cavity
[Y-Y. Liu \textit{et al}, Phys. Rev. Lett. 119, 097702 (2017)].
In this paper, we develop an in-depth, rigorous understanding of this experiment and related works.
%
First, we use a semiclassical theory and study transmission spectroscopy. This approach allows us to
derive the masing threshold condition for arbitrary temperature and voltage bias,
and expose microscopic principles required for realizing photon gain and thereby a photon amplifier.
Next, by employing the quantum master equation approach
we extend the Scully-Lamb quantum theory of a laser to the present setup, 
and investigate the statistics of emitted photons 
below and above the masing threshold 
as a function of experimentally tunable parameters. 
Although our focus is primarily on hybrid quantum dot circuit - quantum electrodynamics systems, 
our approach is adaptable to other light-matter systems where the gain medium consists of a mesoscopic structure.


\end{abstract}

\maketitle


\section{Introduction}
\label{intro}
Photon statistics and other characteristics of light provide critical information for understanding 
fundamental concepts in light-matter interaction 
\cite{singleatom1, RevNori, PNAS1, kontoskondo, kontosnatcom,kontosprx, marco14}
and for developing novel devices \cite{RevNori, kulqda, kulqdb, KLH2015, pettamaserscience,pettamaserprl}.  
Particular systems that recently attracted significant experimental and theoretical interests are 
semiconducting quantum dots integrated with a superconducting qubit architecture 
\cite{Petersson2012, Viennot2013, petta2014, Kulkarni2014, Frey2012,enslinprl,Toida2013,Deng2013, spin-kontos,gpg, tk2011,Mora1,Mora2}. 
These so called quantum dot circuit quantum electrodynamics (QD-cQED) setups offer several advantages over
standard light-matter (radiation field-atom) systems given their
(i) tunability \cite{RevNori}, (ii) scalability \cite{QCL,scale1,scale2} and (iii) versatility \cite{siv1, pd1,pd2, phaselock1}. 
QD-cQED devices combine mesoscopic systems (quantum dots) with quantum optics components. They offer a rich 
platform for the study of light-matter phenomena, as one can investigate both electronic and photonic properties therein. 

Recent remarkable experiments realized parallels to single atom and double-atom masers
in QD-cQED systems \cite{singleatom1, pettamaserscience,pettamaserprl}.
It is to be noted that single atom masers have been previously
realized with Rydberg atoms \cite{Ryd}, optical cavities coupled to either natural or artificial atoms
\cite{art1, art2, art3,art4}, and superconducting junctions \cite{super1,super2}.

Considering QD-cQED systems in a nonequilibrium steady state (NESS),
several complementary quantities can be experimentally observed and theoretically computed.
Recently, in addition to electronic properties, 
the photonic sector has been thoroughly probed \cite{singleatom1,pettamaserscience} 
with measurements reporting on photon transmission, phase response, 
photon number, and the statistics of emitted photons.  
%
The behavior of the electronic degrees of freedom in QD-cQED systems
is examined through the NESS charge current, associated current fluctuations, 
and the quantum dots occupation number \cite{petta2014,Kulkarni2014}.
Altogether, concurrent studies of the photonic and electronic sectors expose effects 
related to light-matter interaction.

The realization of masers in QD-cQED systems calls for a rigorous theoretical description.
Particularly, the observation of masing,  and measurements of 
photon statistics in Double-Quantum-Dot (DQD) masers are missing a careful, 
fundamental quantum  analysis. 
In our previous work \cite{longNEGF}, we investigated the photonic and electronic properties of a DQD setup 
employing the non-equilibrium Green's function (NEGF) approach. 
However, given the perturbative nature of the analysis in the light-matter interaction energy, 
the work was limited to the below-masing threshold regime \cite{longNEGF, QCL}.

\begin{figure}
\begin{center}
\includegraphics[scale=.5]{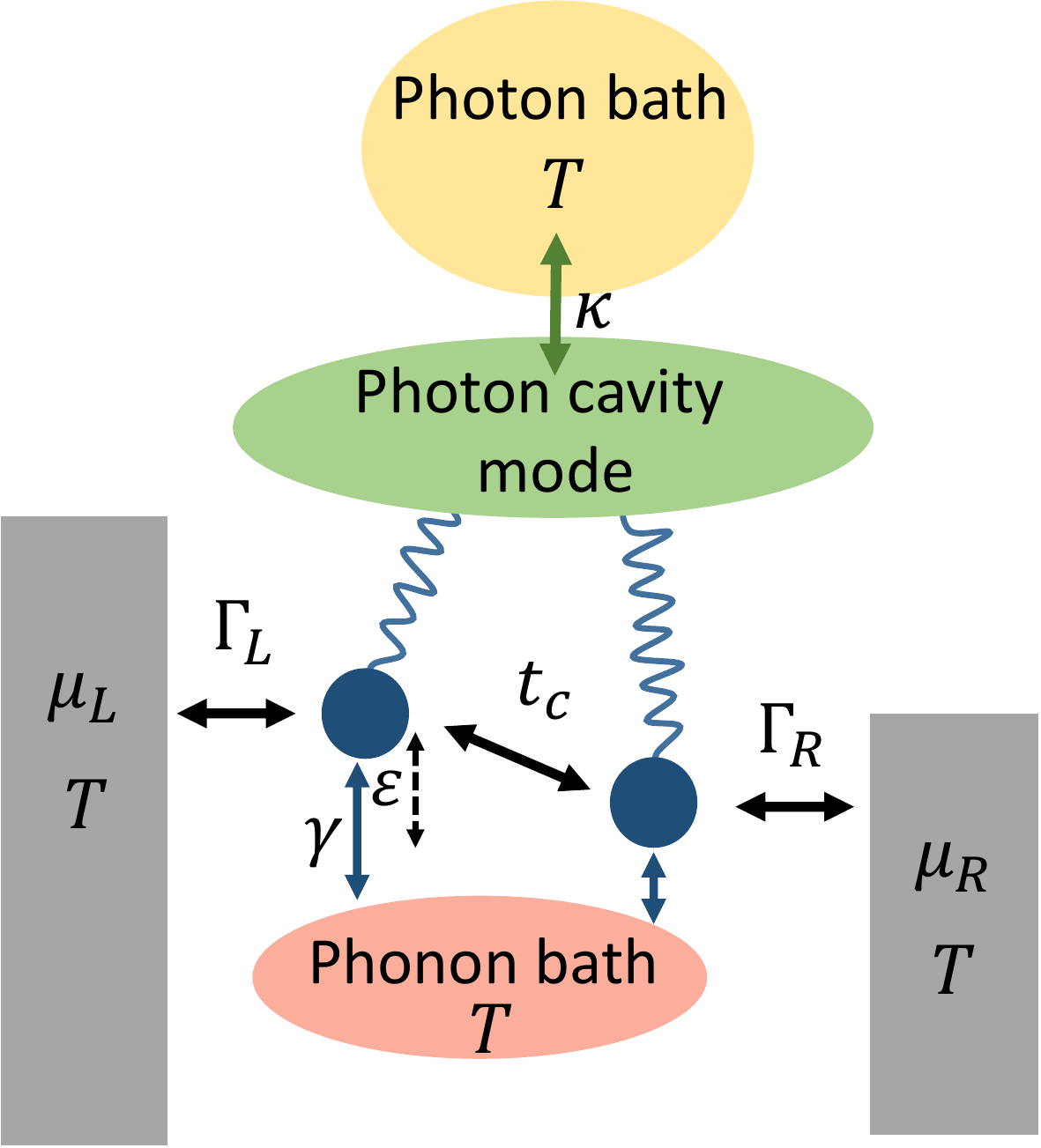}              
\caption{Schematic representation of the model considered in this work.
A double quantum dot (splitting $\epsilon$ and tunneling element $t_c$) 
is bridging metal electrodes (coupling strength $\Gamma_{L,R}$), which are maintained 
at different chemical potentials ($\mu_L$, $\mu_R$) but at
 the same temperature $T$. 
Additionally, the DQD is coupled to a photonic mode, with its own decay channel, and a phononic bath.
We study here the impact of the nonequilibrium electronic medium, that is the voltage-biased 
quantum dot junction on the photonic sector.}
\label{schematic}
\end{center}
\end{figure}

In this paper, we employ the Lindblad quantum master equation (QME) approach and perform a 
careful and comprehensive study of light amplification and masing in a cavity coupled DQD setup, see Fig. \ref{schematic}.
The QME method allows us to investigate, in a unified manner the statistics of photons in the cavity 
as we transit from below to above the masing threshold, unlike the perturbative NEGF approach  \cite{longNEGF}. 
Traditionally, the Lindblad QME approach has been applied to study systems under an 
infinite voltage bias, therefore supporting a unidirectional source-drain current \cite{Kulkarni2014}. 
In contrast, in this paper we extend our investigation to the case of a finite voltage bias, which allows us
to observe the passage of the electronic bath from a dissipative element to a driving, gain medium. 
Given that the source-drain bias can be experimentally tuned, our results, 
capturing voltage-bias dependent physics, shall be useful for understanding and further developing DQD based masers.
The Lindblad QME further allows us to critically examine the detrimental (yet not fatal) impact 
of substrate phonons on photon amplification, considering it in the limit of weak electron-phonon interaction.

We perform transmission spectroscopy under the semiclassical approximation, which limits us to the below-threshold
masing regime. Furthermore, we study photon statistics using a full quantum approach and observe the entire development
of the photon statistics from thermal to Poissonian as we transit from below to above the masing threshold.
The two calculations, transmission spectroscopy and photon statistics, agree on the threshold condition, and they 
provide complementary information on light amplification and masing in our setup.




The paper is organized as follows. 
In section \ref{sec-modelH}, we describe the Hamiltonian of the DQD maser, 
including the external reservoirs (fermionic and bosonic).  
In Section \ref{sec-semi}, based on a semiclassical treatment,
we present analytical results and simulations for photon transmission and phase response  in the system.
In order to rigorously examine the rich nature of photon statistics in the cavity, 
we resort to a fully quantum approach in Section \ref{photon-stat},
%
by developing the quantum theory of lasers due to Scully and Lamb \cite{slbook, Agarwal}.  
We show that the statistics of photons evolves from thermal (with an effective temperature) 
to Poissonian when experimentally-tunable parameters are varied, 
such as bias voltage, level detuning, light-matter coupling, cavity decay rate. 
In Section \ref{sec-summ}, we summarize 
our work and provide an outlook of future challenges. Technical details are delegated to the Appendix.

\section{Model Hamiltonian}
\label{sec-modelH}

We consider an open light-matter quantum system with the total Hamiltonian $\hat H$ consisting of a matter part 
$\hat H_{\rm matter}$ which is driven to an NESS by an external voltage bias, 
a cavity (transmission line resonator) $\hat H_{\rm cavity}$, 
and a light-matter interaction term $\hat H_{\rm matter-cavity}$,
\be
\hat H = \hat H_{\rm matter} + \hat H_{\rm cavity} + \hat H_{\rm matter-cavity}.
\ee
The matter component $\hat H_{\rm matter}$ consists of a double quantum dot placed between 
two fermionic leads maintained at different chemical potentials.
Electron tunneling between the dots takes place via a direct coherent coupling. 
The quantum dots further interact and exchange energy with substrate phonons. 
The matter Hamiltonian therefore consists of the following components,
\be
\hat H_{\rm matter} = \hat H_{\rm DQD} + \hat H_{\rm DQD-lead} + \hat H_{\rm DQD-phonon},
\ee
where the bare dots Hamiltonian is
\bea
\label{HDQD}
\hat H_{\rm  DQD} = \frac{\epsilon}{2} \tau_z + t_c \tau_x.
\eea
The metal leads are included in 
\bea
\hat H_{\rm DQD-lead}&=& 
\sum_{k , \alpha=L,R} \epsilon_{k\alpha} \hat c^{\dagger}_{k \alpha}\,  \hat c_{k \alpha} 
\nonumber \\
&+& \sum_{k} \left[\lambda_{k L} \hat c_{kL} |L\rangle \langle 0| 
+ \lambda_{k R} \hat c_{kR} |R\rangle \langle 0| \right] + {\rm h.c.}, 
\nonumber \\
\eea
and the phononic interaction Hamiltonian is
\bea
\hat H_{\rm  DQD-phonon}&=& \sum_{q} \omega_{q } \hat b^{\dagger}_{q }\,  \hat b_{q } + \tau_z \sum_{ q}\lambda_{ q} 
\big(\hat b_{ q} + \hat b_{q}^{\dagger}\big).
\label{Hmatter}
\eea
Here, $\epsilon$ is the detuning parameter 
and $t_c$ is the direct tunnelling term. 
$\tau_z = |L\rangle \langle L| - |R \rangle \langle R|$ 
and $\tau_x = |L\rangle \langle R| + |R \rangle \langle L|$ represent the $z$ and $x$ components 
of the Pauli matrix, expressed in terms of the localized single electron orbitals, $|L\rangle$ and $|R\rangle$. 
We limit ourselves to the Coulomb blockade regime: This implies that at any instant,
the DQD is restricted to three possible configurations, namely, the null-electron subspace, 
denoted by $|0\rangle$, and the single-electron subspace, with an electron localized either on the left or the right dot, denoted by $|L\rangle$ and $|R\rangle$, respectively. 
We set the energy of the unoccupied electronic state at zero. 

For the DQD-lead Hamiltonian, $\hat c^{\dagger}_{k\alpha} (\hat c_{k \alpha})$ is 
the creation (annihilation) operator for fermions with wave vector $k$ in the $\alpha$-th lead ($\alpha=L,R$). 
$\lambda_{k \alpha}$ is the coupling constant between the DQD and the fermionic bath. 
Similarly, for the DQD-phonon part, $\hat b^{\dagger}_{q} (\hat b_q)$ is the bosonic creation (annihilation) operator 
for phonons with wave vector $q$.  $\lambda_q$ denotes the interaction energy between the DQD and the phonons.  

The photonic component consists of a cavity photon mode of frequency $\omega_c$.
This so-called primary mode (creation operator $\hat a^{\dagger}$)
is coupled to two additional secondary photon baths $(K=L,R)$, which mimic 
the two ends of a long microwave transmission line (creation operator $\hat a_{jK}^{\dagger}$). 
The cavity Hamiltonian is given by
\bea
\hat H_{\rm  cavity} = \omega_c \hat a^{\dagger} \hat a &+& \sum_{j \in K=L,R} \omega_{jK} \hat a_{jK}^{\dagger} 
\hat a_{jK} \nonumber \\ &+& \sum_{j \in K=L,R} \nu_{j} \hat a_{jK}^{\dagger} \hat a + {\rm h.c.}.
\label{cavity}
\eea
Finally, light (cavity)-matter(DQD) interaction is given by the standard dipole coupling term,
\be
\hat H_{\rm matter-cavity}= g \, \tau_z \,(\hat a + \hat a^{\dagger}),
\label{matter-cavity}
\ee
where $g$ is the light-matter coupling constant. 
This model is extremely compound---and rich---offering different regimes of operation (weak/strong coupling
of the DQD to the leads, phonons, cavity mode, weak/strong dissipation of the cavity mode to secondary modes, 
linear response/far from equilibrium operation).

In what follows, we first diagonalize the DQD Hamiltonian in Eq.~(\ref{HDQD}) and work in the diagonalized basis. 
This new eigenbasis relates to the localized states $|L\rangle, |R\rangle$ via an unitary transformation, 
\bea
|g\rangle &=& \cos\left(\frac{\theta}{2}\right) |L\rangle + \sin\left(\frac{\theta}{2}\right) |R\rangle, \nonumber \\
|e\rangle &=& -\sin\left(\frac{\theta}{2}\right) |L\rangle +\cos\left(\frac{\theta}{2}\right) |R\rangle.
\label{transformation}
\eea
Here, $|g\rangle,|e\rangle$ represents the ground and excited states for the DQD,
 and $\theta= \arctan (-2t_c/\epsilon)$. In the energy basis, the full Hamiltonian transforms to 
%
\be
\hat H_{\rm DQD}= \frac{\Omega}{2} \big(\hat d_{e}^{\dagger} \hat d_e - \hat d_{g}^{\dagger} \hat d_g\big),
\ee
and
\bea
&&\hat H_{\rm DQD-lead} = \nonumber \\
&&\sum_{k} 
\begin{pmatrix} t_{k L} \hat c_{k L} &  t_{k R} \hat c_{k R}\end{pmatrix}
  \begin{pmatrix} \cos\left(\frac{\theta}{2}\right) &  -\sin\left(\frac{\theta}{2}\right) \\
 \sin\left(\frac{\theta}{2}\right) & \cos\left(\frac{\theta}{2}\right) \end{pmatrix} 
 \begin{pmatrix} \hat d_e^{\dagger} \\ \hat d_g^{\dagger} \end{pmatrix} \nonumber \\
&+& {\rm h.c.}, 
\eea
\bea
\hat H_{\rm DQD-phonon}&=& \sum_{q} \omega_{q}  \hat b^{\dagger}_{q}\,  \hat b_{ q} 
+\sum_{q}\lambda_{q} \Big[ \cos\theta \big(\hat d_e^{\dagger} \hat d_e - \hat d_g^{\dagger} \hat d_g\big)
\nonumber \\
&&- \sin\theta \big(\hat d_e^{\dagger} \hat d_g + \hat d_g^{\dagger} \hat d_e \big)\Big] \big(  \hat b_{q} + \hat b_{q}^{\dagger}\big), \\
\hat H_{\rm matter-cavity}&=& -g \sin\theta (\hat d_e^{\dagger} \hat d_g \hat a + \hat d_g^{\dagger} \hat d_e  \hat a^{\dagger}).
\label{eq:mattercavity}
\eea
Here, $\hat d_{e,(g)}^{\dagger} = |e\rangle \langle 0|,  (|g\rangle \langle 0|)$ is the creation operator for the
 excited, (ground) state and $\Omega= \sqrt{\epsilon^2 + 4 t_c^2}$ is the DQD renormalized frequency.  
We also define the eigenenergies of the DQD as $\epsilon_g=-\Omega/2$ and $\epsilon_e=\Omega/2$. 
To arrive at this matter-cavity Hamiltonian, we perform the rotating wave approximation (RWA) 
assuming that the cavity frequency $\omega_c$ is in resonance with the DQD energy gap $\Omega$. 
However, such RWA is not justified for the DQD-phonon Hamiltonian since the frequency spectrum 
of the phonon bath can be off-resonant with the DQD energy gap. 
Note that the cavity Hamiltonian in Eq.~(\ref{cavity}) is not modified under this transformation. 

In what follows, considering this DQD setup, we first discuss photon transmission spectroscopy 
using a semi-classical approach. This discussion is followed by a quantum treatment for photon statistics. 

\section{Semiclassical theory of Photon transmission}
\label{sec-semi}
\subsection{Threshold condition for masing} 

We first investigate the cavity response using transmission spectroscopy, calculating
the transmission amplitude and the phase response for emitted microwave photons. 
In transmission spectroscopy measurements, the cavity is driven with a coherent microwave field;
the output field is measured via heterodyne detection \cite{petta2014}. 
In this section we assume that light-matter interaction is weak, and that the driving field is weak.
The resulting QME then correspond to the semiclassical  limit, as we explain below.

We employ the quantum master equation approach to derive an expression for the transmission. 
We write down an equation of motion for the reduced density matrix for the DQD + cavity mode, 
$\rho={\rm Tr}_{\rm reservoirs}\big[\rho_{\rm total}\big]$ by tracing out all reservoirs' 
degrees of freedom, that is the electronic, photonic and phononic reservoirs. 
We make use of the standard set of approximations, namely, 
the system-bath decoupled initial condition for the density matrix, 
the Markov approximation, and a weak-coupling treatment between the system and the different reservoirs. Similar approximations have been used in other studies of QD-cQED systems when exploring
full-counting statistics, transport and photonic properties 
\cite{Vavilov-fcs,Schon_noise,ss1,ss2,Nori-bistable,siv2,Schon, Child, Jin, Muller}. We then
arrive at the following equation,
\be
\dot{\rho}= -i [\hat H_0(t), \rho] + {\cal L}_{\rm electron}[\rho] + {\cal L}_{\rm photon}[\rho] + {\cal L}_{\rm phonon}[\rho].
\ee
The first term represents the quantum coherent time evolution governed by the Hamiltonian 
$\hat H_0(t)=\hat H_{\rm DQD} + \hat H_{\rm matter-cavity} + \omega_c \hat a^{\dagger} \hat a + i \sqrt{\frac{\kappa}{2}} E \cos(\omega_d t) (\hat a^{\dagger} -\hat a)$. 
The last term in $\hat H_0(t)$ is a coherent driving term,
which represents the transmission measurement \cite{petta2014,Kulkarni2014}.
The last three terms in the QME correspond to different Liouvillians 
capturing the effects of the electronic, photonic, and phononic reservoirs, respectively.
The electronic Liouvillian collects the effect of the two metals on the system,
\be
{\cal L}_{\rm electron}[\rho]= \sum_{{\alpha=L,R},\\{n=e,g}} {\cal L}_{\alpha n}[\rho]
\ee
where 
\be
{\cal L}_{\alpha n} [\rho] 
= \frac{1}{2} \Gamma_{\alpha n} (\theta) \Big[ f_{\alpha}(\epsilon_n) D[\hat d_n, \rho] +  (1-f_{\alpha}(\epsilon_n)) D[\hat d^{\dagger}_n, \rho]\Big].
\ee
Here,  
\bea
\Gamma_{Le(Rg)}(\theta) &=& \Gamma_{Le(Rg)} \cos^2\left(\frac{\theta}{2}\right), 
\nonumber \\
\Gamma_{Lg(Re)}(\theta) &=& \Gamma_{Lg(Re)} \sin^2\left(\frac{\theta}{2}\right).
\label{eq:Ga}
\eea 
$\Gamma_{\alpha n}$ is the spectral function  (or hybridization) of the electronic lead $\alpha$ with the state $n$.
For simplicity, it is chosen to be flat (wide-band limit). 
$f_{\alpha}(\epsilon_n)= [\exp{(\beta(\epsilon_{n}-\mu_{\alpha}))} +1]^{-1}$ is the Fermi distribution for the lead $\alpha$ with chemical potential $\mu_{\alpha}$ and inverse temperature $\beta=1/k_B T$. 
The dissipator $D[\hat O, \rho]$ for an operator $\hat O$ is defined as
\be
D[\hat O, \rho] = 2\, \hat O^{\dagger} \,\rho\, \hat O - \{\hat O \, \hat O^{\dagger}, \rho\}.
\ee
We can similarly write down expressions for ${\cal L}_{\rm photon} [\rho]$     
and ${\cal L}_{\rm phonon} [\rho]$. 
Assuming the cavity is in contact with a very low temperature transmission line, the photonic Liouvillian is
\be
{\cal L}_{\rm photon} [\rho] = \frac{\kappa}{2} D[\hat a^{\dagger}, \rho],
\ee
%
where $\kappa_{K}=2 \pi F_{K} |\nu|^2$  
is the cavity decay rate per port ($K=L,R$), and $\kappa= \kappa_L + \kappa_R$ 
is the total decay rate. Here $F_{K}$ is the density of states of the $K$-th photonic bath and $\nu$ is the average coupling between the cavity and photon bath modes. 
For simplicity, we assume a symmetric decay rate i.e., $\kappa_L=\kappa_R = \kappa/2$. 
Lastly, the Liouvillian due to the phonon environment is given as
\bea
{\cal L}_{\rm phonon}[\rho]&=& \frac{\gamma_{u}(\Omega)}{2} D[\hat d_g^{\dagger} \hat d_e, \rho] 
+ \frac{\gamma_{d}(\Omega)}{2} D[\hat d_e^{\dagger} \hat d_g , \rho] + \nonumber \\
&&\frac{\gamma_{\phi}{(0)}}{2} D[(\hat d_e^{\dagger} \hat d_e - \hat d_g^{\dagger} \hat d_g), \rho].
\eea
Here, $\gamma_{u}$, $\gamma_{d}$ and $\gamma_{\phi}$ are the phonon pumping, relaxation and pure dephasing rate constants, 
$\gamma_{u}(\Omega) = 2 \sin^2(\theta) \, n_{th}(\Omega)\, J(\Omega)$, 
$\gamma_{d}(\Omega) = 2 \sin^2(\theta) \, \left[1+ n_{\rm th}(\Omega)\right] \, J(\Omega)$ 
and $\gamma_{\phi}(0) = 2 \cos^2(\theta) \, [1+ 2 n_{th}(0)]\,J(0)$.
Recall that the DQD frequency is defined as $\Omega=\epsilon_{e}-\epsilon_g$.
The phononic rate constants are given in terms of $n_{\rm th}(\omega)= 1/(e^{\beta \hbar \omega}-1)$,  
the Bose-distribution function with inverse temperature    
$\beta= 1/k_B T$, and $J(\omega)$, which is the phonon spectral function. It is
chosen to be of the following form \cite{petta2014, Kulkarni2014, petta-phonon, weber-phonon}, 
%
%
\be
J(\omega) = j_{\rm piezo}\, \left(\frac{\omega}{\omega_0}\right)\, e^{-\omega^2/\omega_D^2}\, \Big[1- \sin\Big(\frac{\omega}{\omega_D}\Big)\Big],
\ee 
where $\omega_D$ and $\omega_0$ are the scaling parameters, $j_{\rm piezo}$ is the coupling strength. The typical parameter values are given in our Table.~1.

We now write down equations of motion for the population and coherences of the DQD states 
and the photon mode under the semiclassical approximation, 
given by,  $\langle \hat d_e^{\dagger} \hat d_g \hat a \rangle \approx \langle \hat d_e^{\dagger} \hat d_g \rangle \langle \hat a \rangle$. 
We receive,
\begin{widetext}
\bea
\dot{\rho}_{ee} &=& \Big(\Gamma_{Le}^c + \Gamma_{Re}^s\Big) \rho_{00} - \Big(\bar{\Gamma}_{Le}^c + \bar{\Gamma}_{Re}^s + \gamma_{d} \Big) \rho_{ee} + \gamma_{u} \rho_{gg} + i g \sin\theta \Big( \rho_{ge} \langle \hat a \rangle  - h.c\Big), \\
\dot{\rho}_{gg} &=& \Big(\Gamma_{Lg}^s + \Gamma_{Rg}^c\Big) \rho_{00} - \Big(\bar{\Gamma}_{Lg}^s + \bar{\Gamma}_{Rg}^c + \gamma_{u} \Big) \rho_{gg} + \gamma_{d} \rho_{ee} - i g \sin\theta \Big( \rho_{ge} \langle \hat a \rangle  - h.c\Big), \\
\dot{\rho}_{eg} &=& - i \Omega \rho_{eg} - i g \sin \theta \Big(\rho_{ee} \!-\! \rho_{gg}\Big) \langle \hat a \rangle - \Big( \frac{1}{2} \Gamma_{\rm eff} + 2 \gamma_{\phi}\Big) \rho_{eg}, \\
{\langle \dot {\hat{a}} \rangle} &=& -i \omega_c \langle \hat a \rangle -\frac{1}{2} \Big( \kappa \langle \hat a \rangle - 2 i g \sin\theta \rho_{eg} \Big) + \sqrt{\frac{\kappa}{2}} E \cos(\omega_d t).
\label{eq-master}
\eea
\end{widetext}
Here we use the following compact notation, 
$\Gamma_{Le(Rg)}^c\!=\!\Gamma_{Le (Rg)}(\theta) f_{L(R)}(\epsilon_{e(g)})$,
$\Gamma_{Lg (Re)}^s\!=\! \Gamma_{Lg (Re)}(\theta) f_{L(R)}(\epsilon_{g(e)})$; the superscripts follow from the sin (s) and 
cos (c) functions  in Eq. (\ref{eq:Ga}).
The bar symbol replaces the Fermi function $f$ by $1\!-\!f$. 
For example, $\bar{\Gamma}_{Le (Rg)}^c\!=\! {\Gamma}_{Le (Rg)}(\theta) \big[1\!-\!f_{L(R)}(\epsilon_{e(g)})\big]$.
$\Gamma_{\rm eff} = \Big( \bar{\Gamma}_{Lg}^s + \bar{\Gamma}_{Rg}^c + \bar{\Gamma}_{Le}^c + \bar{\Gamma}_{Re}^s + \gamma_{u} + \gamma_{d}\Big)$
is the effective damping constant and is a function of the different tuning parameters of the electronic medium (DQD+leads)
such as the chemical potentials, temperature of the electronic leads, coupling energy between the DQD and the
electronic and phononic environments.

We next solve the above set of equations in steady state by moving to a rotating frame with respect to the 
driving frequency $\omega_d$ and assuming 
$\langle \hat a \rangle(t) \approx \langle \hat a \rangle_{ss} e^{-i \omega_d t}$ 
and $\rho_{eg}(t) \approx \rho^{ss}_{eg}\,e^{-i \omega_d t}$. We get
\be
\rho_{eg}^{ss}= \frac{g \sin(\theta) (\rho_{gg}-\rho_{ee})}{(\omega_d-\Omega) + i \big(\frac{1}{2} \Gamma_{\rm eff} + 2 \gamma_{\phi}\big)} \langle \hat a \rangle_{ss}.
\ee
Substituting this expression into the equation of motion $\langle \dot {\hat a}\rangle $  in Eq. (\ref{eq-master}) 
and solving it in steady state we obtain the
transmission function,
\be
t(\omega_d) 
\equiv  \frac{\sqrt{2 \kappa} \langle \hat a \rangle_{ss}}{E} =\frac{i \kappa/2}{(\omega_d-\omega_c) + i \kappa/2 - \chi_{\rm el}(\omega_d)},
\label{trans0}
\ee
where we identify $\chi_{el}(\omega)$ as the charge susceptibility, given as 
\be
\chi_{\rm el}(\omega) = \frac{g^2 \sin^2(\theta)}{(\omega-\Omega) +i \big(\frac{1}{2} \Gamma_{\rm eff} + 2 \gamma_{\phi}\big)} (\rho^{ss}_{gg}\!-\!\rho^{ss}_{ee})|_{g=0}.
\ee
The population of the DQD states, $\rho_{gg}^{ss},\rho_{ee}^{ss}$, 
are evaluated in the nonequilibrium steady state and in absence of light-matter interaction.  
The susceptibility, which contain information about population imbalance, has a non-trivial dependence 
on experimentally tunable parameters. 
Note that arriving at Eq.~(\ref{trans0}) following Eq.~(\ref{eq-master}) is valid only in the limit 
\bea
\kappa &\geq& 2 \, {\rm Im} \big[\chi_{el}(\omega_d)\big] \nonumber \\
&=& \frac{-2 g^2 \sin^2(\theta) \, \big(\frac{1}{2} \Gamma_{eff} + 2 \gamma_{\phi}\big)}{\big(\omega_d-\Omega)^2 + \big(\frac{1}{2} \Gamma_{\rm eff} + 2 \gamma_{\phi}\big)^2} \big(\rho_{gg}^{ss}  - \rho_{ee}^{ss}\big)|_{g=0}.
\label{threshold-semi}
\eea
We identify the right hand side of this inequality as the masing threshold.  
When using the NEGF formalism, a masing threshold limit for a similar model had been 
derived in our previous work \cite{longNEGF}, based on the causality condition for the Green's function.  

We rewrite the transmission as $t(\omega_d) = |t(\omega_d)| e^{i \phi(\omega_d)}$;
 gain in the cavity photon is  $|t(\omega_d)|>1$ and the phase response is included in $\phi(\omega_d)$. 
It is immediately clear from Eq.~(\ref{trans0}) that to achieve photon gain 
one has to counteract two different sources of dissipation: the cavity decay to the ports  (rate constant $\kappa$)
and the imaginary component 
of the electronic medium induced charge susceptibility. 
At equilibrium, $\rho_{gg} > \rho_{ee}$, ${\rm Im}[\chi_{\rm el}(\omega)] <0$, 
which immediately implies that photon gain is impossible to achieve in this limit:
At equilibrium, in addition to the photon bath, the electronic degrees of freedom
further act as a dissipative channel for cavity photons. 

It is only when the DQD is driven far from equilibrium, 
population inversion happens,  ${\rm Im}[\chi_{\rm el}(\omega)]$ changes sign,
and photon gain is achieved. 
Therefore, by driving the DQD out of equilibrium one can realize a photon amplifier. 
More interestingly, one can further derive a sum rule for the transmission function, 
following the definition in Eq.~(\ref{trans0}), given as
\be
\int_{-\infty}^{\infty} \frac{d\omega_d}{2 \pi} \, t(\omega_d) = \frac{i \kappa}{4},
\label{Eq:sum-rule}
\ee
valid in the regime $\kappa \geq 2 \, {\rm Im} \big[\chi_{el}(\omega_d)\big]$, as mentioned before.
The general analytical expression for the transmission function, the threshold condition and the sum rule 
are the first set of central results of this paper.  



\begin{center} 
Table 1: Typical parameter values from experiments (Refs.~\onlinecite{petta2014,Kulkarni2014})
\begin{tabularx}{\columnwidth}{X l l  }
\hline
\hline
Cavity loss rate $\kappa$ & $0.0082$ $\mu$eV&$2.0$ MHz\\
Light-matter coupling $g$& $0.2050$ $\mu$eV& $50$ MHz\\
Cavity frequency $\omega_c$&  $32.5$ $\mu$eV&$7.86$ GHz  \\ 
Elastic tunneling $t_c$ & $16.4$ $\mu$eV& $3.96$ GHz\\
detuning $\epsilon$ & $20$ $\mu$eV&  $4.84 $ GHz \\
Drain tunneling rate $\Gamma_R$ & $16.56$ $\mu$eV&$4.0$ GHz \\ 
Source tunneling rate $\Gamma_L$ & $16.56$ $\mu$eV&$4.0$ GHz \\
$j_{piezo}$ (unknown)  & $5.96$ $\mu$eV& $1.44$ GHz \\ 
Scaling frequency $\omega_0$ &$32.8$ $\mu$eV&$7.9$ GHz  \\
Phonon bath cutoff $\omega_D$  & 35 $\mu$eV & $8.46$ GHz \\ 
Temperature $T$ & $8$ mK& 0.16 GHz \\
\hline
\hline
\label{tab:title}
\end{tabularx}\par
\end{center}


\begin{figure}
\begin{center}
\includegraphics[height=4cm, width=\columnwidth, scale=1.2]{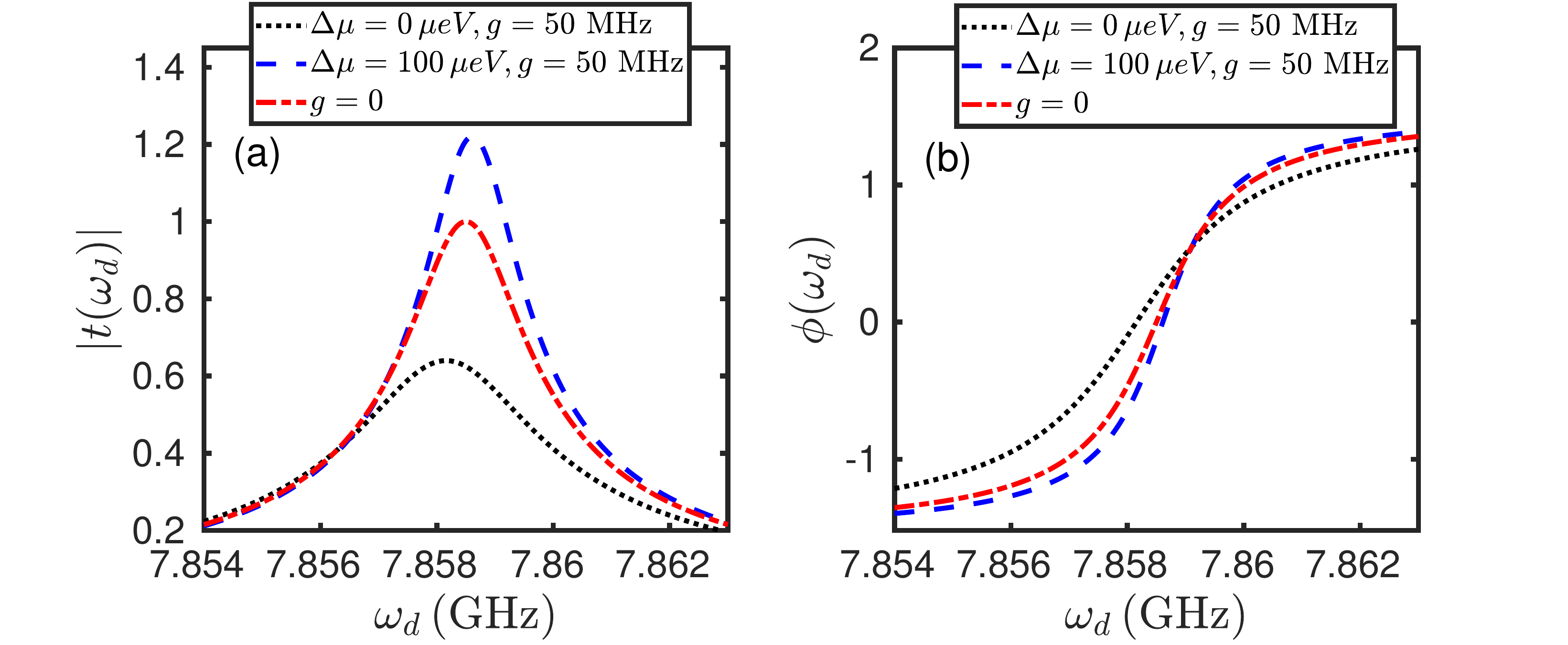}     
\caption{Transmission spectroscopy for cavity photon: 
(a) absolute value of transmission $|t(\omega_d)|$ and 
(b) phase $\phi(\omega_d)$ as a function of the incoming frequency $\omega_d$
in the absence of the cavity-matter coupling $g=0$ MHz (red, dashed-dotted), 
in the presence of the DQD, which is maintained at equilibrium $\Delta \mu=0$ (black, dotted), 
or driven out of equilibrium $\Delta \mu$=100 $\mu eV$ (blue, dashed). 
Parameters are $\epsilon$=20 \,$\mu eV$, $t_c$=16.4 \, $\mu eV$; other parameters are reported in Table~1.}
\label{trans_omega}
\end{center}
\end{figure}

\begin{figure}
\includegraphics[scale=0.7]{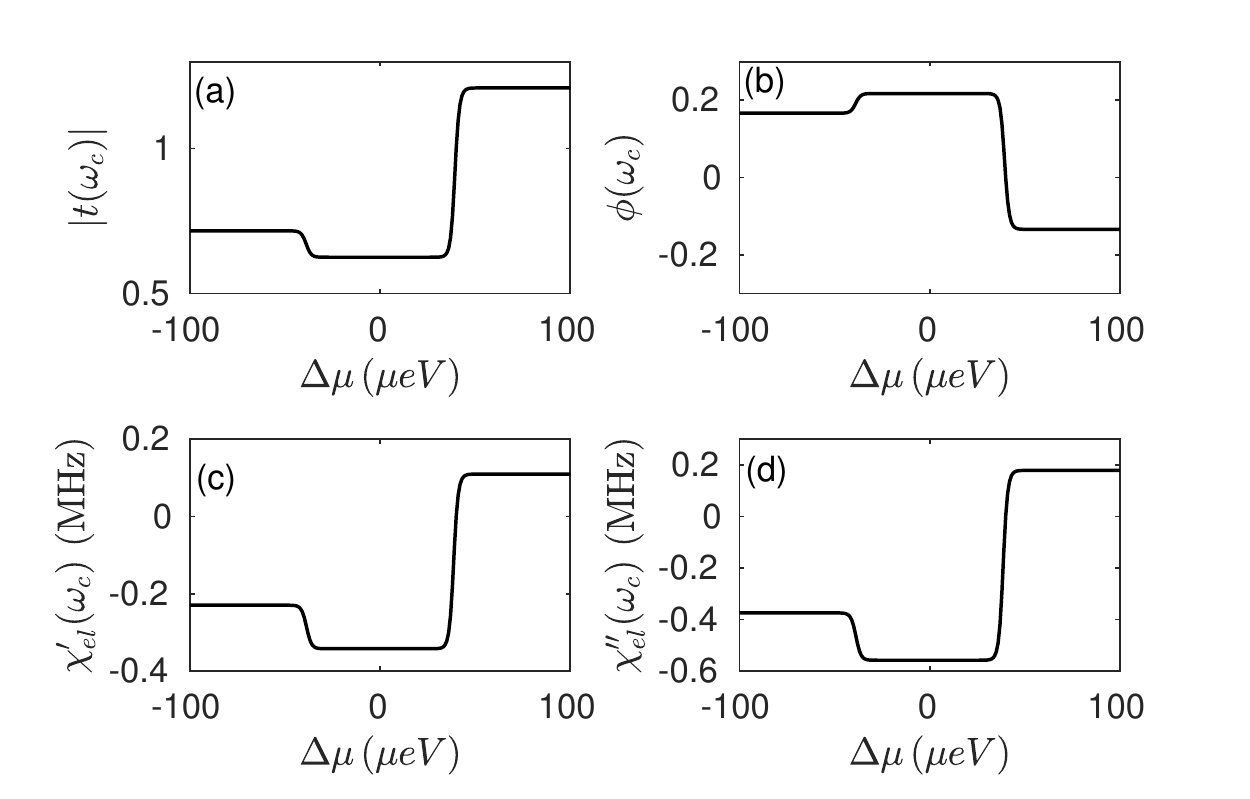}               
\caption{
(a) Absolute value of transmission $|t(\omega_c)|$, 
(b) phase $\phi(\omega_c)$, 
(c) real part of the charge susceptibility, $\chi^{'}_{\rm el}(\omega_c)$, 
and (d)  imaginary part of the charge susceptibility, $\chi^{''}_{\rm el}(\omega_c)$. 
Calculations are performed
 in the presence of phonons as a function of the external bias voltage $\Delta \mu$. 
Here $\epsilon$ =20 $\mu eV$, $\Delta \mu$ =100 $\mu eV$,  with other parameters reported in Table~1.}
\label{trans_bias}
\end{figure}

\begin{figure}
\includegraphics[scale=0.7]{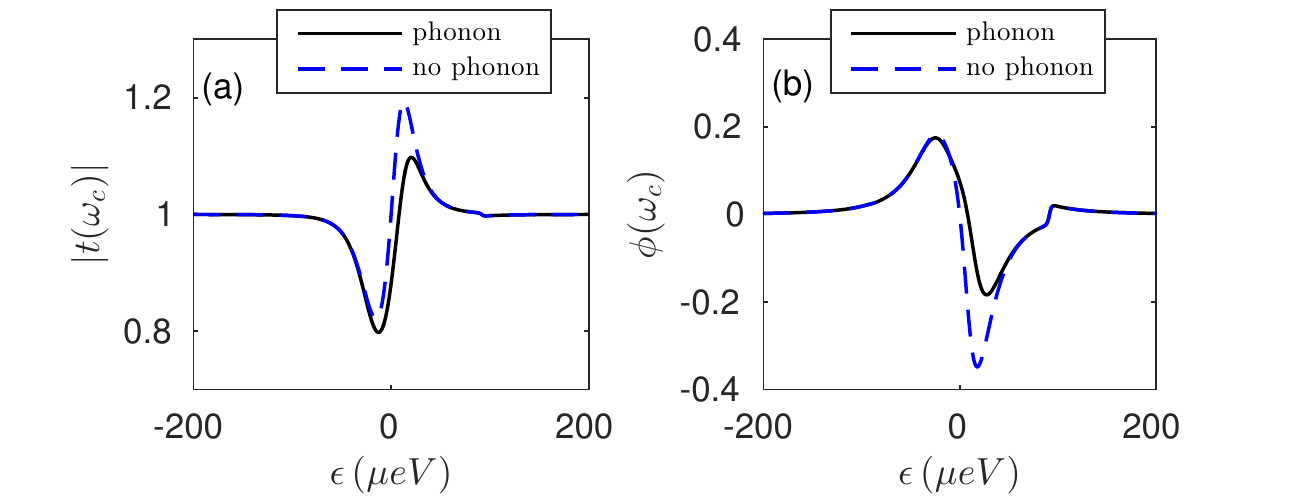}        
\caption{ 
(a) Absolute value of transmission $|t(\omega_c)|$ and (b) phase $\phi(\omega_c)$ at the cavity frequency $\omega_c$ 
as a function of the detuning $\epsilon$. 
The parameters are $t_c=20$ $\mu eV$, $\Delta \mu =100 \mu eV$; other parameters are reported in Table~1.}
\label{trans_detu}
\end{figure}

\begin{figure}
\begin{center}
\includegraphics[scale=0.6]{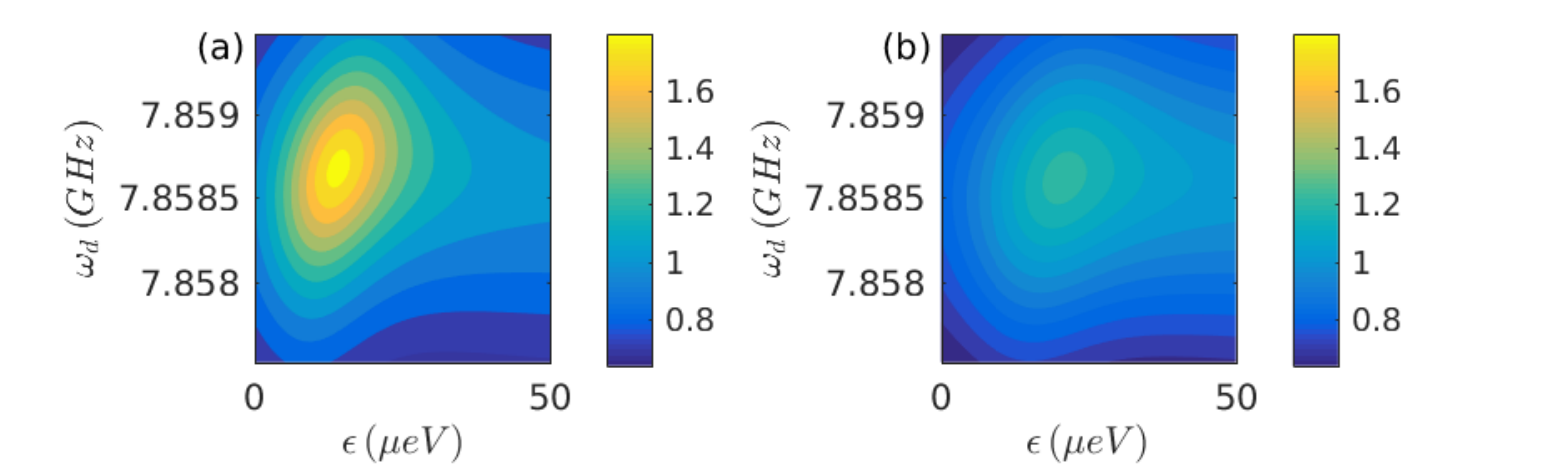}  
\caption{Contour plot of the absolute value of transmission $|t(\omega_d)|$ as a function of 
incoming frequency $\omega_d$ and detuning $\epsilon$
(a) without and (b) with substrate phonons. 
Here, $t_c=16.4$ $\mu$eV, $\Delta \mu$=200 $\mu$eV, with
other parameters reported in  Table~1.}
\label{trans_detu_contour}
\end{center}
\end{figure}

\subsection{Numerical results for photon transmission}

We present numerical results for photon transmission as a function of 
experimentally tunable variables: voltage bias, DQD parameters,
driving frequency; values for relevant parameters are given in Table~1. 
Unless otherwise stated, we set the Fermi energy of the electronic leads at zero, and 
symmetrically adjust the voltage around it,  $\mu_L = -\mu_R = \Delta \mu /2$. 
We set the temperature of all the baths to be the same. 
For the spectral function of the electronic leads we use the wideband approximation 
and choose symmetric couplings, $\Gamma_L = \Gamma_R = \Gamma$. 
 
In Fig.~\ref{trans_omega}(a) we plot the absolute value of the photon transmission $|t(\omega_d)|$
as a function of the incoming coherent microwave frequency $\omega_d$.
The corresponding phase $\phi(\omega_d)$ is displayed in Fig.~\ref{trans_omega}(b).  
We observed the following:

First, in the absence of the matter-cavity interaction, the transmission is exactly unity at 
the cavity frequency $\omega_d=\omega_c$ 
and it displays a broadening proportional to $\kappa$ (dashed line).   
Correspondingly, the phase response is zero at the resonant frequency $\omega_c$ and asymptotically it reaches 
$\pm \pi/2$ in the off-resonant regime ($\omega_d \gg \omega_c$). 

Once the cavity-matter interaction is switched on---yet keeping all the baths at equilibrium 
with the same temperature  and chemical potentials---the maximum value of the transmission drops below unity 
with corresponding frequency value shifting from the bare cavity frequency $\omega_c$ (dotted line). 
This shift is due to charge fluctuations in the dots, 
and it is directly proportional to the real part of the charge susceptibility ${\rm Re}[\chi_{\rm el}(\omega)]$. 
The broadening of the transmission function is related to the 
difference between $\kappa$ and ${\rm Im}[\chi_{\rm el}(\omega)]$;
recall that the latter is negative when it acts as a dissipative medium.  
Since at equilibrium ${\rm Im}[\chi_{\rm el}(\omega)]<0$, the broadening is large. 
In other words, the increased broadening at equilibrium compared to the isolated cavity-matter case
implies that the electronic component acts as a dissipative channel for the cavity photons. 
It is easy to note that the phase response is zero when the transmission is maximal.

Next, the DQD is voltage biased. Once it is driven sufficiently far from equilibrium 
(here $\Delta \mu =100 \, \mu eV > \omega_c$), the absolute value of the transmission exceeds unity (dashed line).
This enhancement is accompanied by a reduction in broadening 
as ${\rm Im}[\chi_{\rm el}(\omega)]>0$ due to population inversion in the DQD states. 
This can also be understood from the sum rule formula in Eq.~(\ref{Eq:sum-rule}), 
which indicates that a reduction in broadening leads to an enhancement in the peak value of the transmission.

To better understand the photon signal, we display in Fig.~\ref{trans_bias} the transmission amplitude, phase, 
and the real and imaginary components of the charge susceptibility, all as a function 
of bias voltage $\Delta \mu$ at the (fixed) cavity frequency $\omega_c$. 
In the close to equilibrium regime, $\Delta \mu < \omega_c$, the absolute value of the transmission is less than unity, 
as expected, while it shows an enhancement for higher bias $\Delta \mu > \omega_c$. 
This shows as a sudden dip in the phase. 
The sudden jump takes place when both the real and imaginary components of the $\chi[\omega_c]$ change sign. 
In particular, when $\chi''(\omega_c)$ becomes positive, photon gain is observed. 

In Fig.~\ref{trans_detu} we plot the photon signal at the cavity frequency as a function of 
the detuning $\epsilon$ of the DQD for a large bias voltage. 
Both absolute value of transmission and the phase display  the peak $(|t(\omega_c)|>1)$ and dip $(|t(\omega_c)|<1)$ structure. 
In the transmission amplitude, for positive detuning $\epsilon>0$ (negative detuning $\epsilon<0$), 
electron transport through the DQD is assisted by light, reaching resonance condition via the  
emission (absorption) of photons and thereby reflected as a peak (dip). 
At large detuning, the cavity and matter units effectively decouple and the transmission value settles to unity with zero phase response.

The unfavorable role of the phonon environment is illustrated in
Fig.~\ref{trans_detu_contour}.  Here, we display a contour plot for the absolute value of transmission as a function of 
incoming frequency $\omega_d$ and detuning $\epsilon$ in the  absence (a) or presence (b) of phonons.
Ignoring the substrate phonons reduces the broadening and results in a further gain in the photon signal. 
Nevertheless, we find that the transmission can exceed unity even under the dissipative action of phonons.
It should be reminded that in this work electron-phonon interaction is assumed weak; Lindblad dissipators agree with
treatments based on second order system-bath perturbation theory. Specifically, 
the additive nature of the Lindblad dissipators in the different baths reflects the absence of bath-cooperative effects.
As such, phonons are detrimental to photon gain since they assist in the dissipation of electronic excitations within the DQD.
In contrast, when cooperative photon-electron-phonon processes are realized (through strong coupling interactions), 
phonon-assisted gain, beyond the strict resonance condition, $\Omega = \omega_c$, is achieved \cite{pettamaserprl}. 

So far, we have considered the semiclassical limit. We acquired the threshold condition
and analyzed photon amplification below threshold.
However, a full quantum approach is required to understand the properties in the above-threshold regime, 
which is related to the masing phenomenon. 
We address this issue in the next Sec.~\ref{photon-stat}.


\section{Quantum Theory of Photon Statistics: Scully-Lamb approach} 
\label{photon-stat}

In this section, we focus on the statistics of the cavity mode and follow its behavior as we transit from below to above
the masing threshold. 
We define the reduced density operator for the cavity photon (ph) as
$\rho_{\rm ph}(t)= {\rm Tr}_{\rm el+phonon + photon-bath}[ \rho(t)]$ and investigate its 
population dynamics, $p_m(t) = \langle m | \rho_{\rm ph}(t) | m \rangle$. 
Following the standard QME procedure, as done before, we obtain
%
\bea
\frac{d}{dt} p_{m} &=& 
i g \sin \theta \Big[ \sqrt{m+1} \big( \rho_{ge; m+1,m} - \rho_{eg; m,m+1}\big)
\nonumber \\
&+& \sqrt{m} \big( \rho_{eg; m-1, m} - \rho_{ge; m,m-1}\big)\Big] 
\nonumber \\
&+& \kappa(1+ \bar{n}) \big[(m+1) p_{m+1} - m p_m\big] 
\nonumber \\
&+&\kappa \bar{n} \big[ m p_{m-1}- (m+1) p_m \big]. 
\label{photon-pop}
\eea
%
The first term (explicit $g$ dependence)  describes coherent evolution and
it consists of joint cavity and DQD density matrix elements. The latter part (explicit $\kappa$ dependence) 
is due to the interaction of the 
cavity mode with the transmission line (photon bath) and it is responsible for the decay of cavity photons
 with rate $\kappa$. 
Here $\bar{n} = \big[\exp({\beta \hbar \omega_c})-1\big]^{-1}$ is the Bose-Einstein distribution function 
of the photon mode at the photon bath temperature $T=1/(k_B \beta)$ and frequency $\omega_c$. 

In order to close equation (\ref{photon-pop}), we need to express 
the joint cavity - DQD density matrix elements
$\rho_{ge;m+1,m}$ and $\rho_{eg;m,m+1}$ in terms of the cavity mode populations $p_m$.
To achieve that, we write down equations of motion for these elements, then
make a crucial approximation that the DQD relaxes to the nonequilibrium steady state 
much faster than the cavity mode, which is indeed the case within our parameters, $\kappa$, $g$ $\ll \Gamma$.  
The equations for these combined density matrix elements are  
%
\begin{widetext}
\bea
\dot{\rho}_{gg;m,n} =&& - i \omega_c (m-n) {\rho}_{gg;m,n} + i g \sin\theta \Big[\sqrt{m} \rho_{eg; m-1,n} -\sqrt{n} \rho_{ge; m,n-1}\Big] + 
\left(\Gamma_{Lg}^s + \Gamma_{Rg}^c \right) \rho_{00;\,m,n} \nonumber \\
&& - \left(\bar{\Gamma}_{Lg}^s + \bar{\Gamma}_{Rg}^c \right) \rho_{gg;m,n} - \gamma_{u} \, \rho_{gg;m,n} + \gamma_{d} \,\rho_{ee;m,n},
\label{eq:r1}
\eea
\bea
\dot{\rho}_{ee;m-1,n-1} =&& - i \omega_c (m-n) {\rho}_{ee;m-1,n-1}+ i g \sin\theta \Big[\sqrt{m} \rho_{ge; m,n-1} -\sqrt{n} \rho_{eg; m-1,n}\Big] + 
\left(\Gamma_{Le}^c + \Gamma_{Re}^s \right) \rho_{00;\,m-1,n-1} \nonumber \\
&& - \left(\bar{\Gamma}_{Le}^c + \bar{\Gamma}_{Re}^s \right) \rho_{ee;m-1,n-1} + \gamma_{u} \, \rho_{gg;m-1,n-1} - \gamma_{d} \,\rho_{ee;m-1,n-1},
\label{eq:r2}
\eea

\bea
\dot{\rho}_{eg;m-1, n} =&& - i \omega_c (m\!-\!n\!-\!1) {\rho}_{eg;m-1,n} -i \Omega {\rho}_{eg;m-1, n} + i g \sin\theta \Big[\sqrt{m} \rho_{gg; m,n} -\sqrt{n} \rho_{ee; m-1,n-1}\Big] \nonumber \\
&&- \left(\frac{1}{2} \Gamma_{\rm eff} + 2 \gamma_{\phi} \right) \, {\rho}_{eg;m-1,n}, 
\label{eq:r3}
\eea
\bea
\dot{\rho}_{ge;m,n-1} =&& - i \omega_c (m\!-\!n+1){\rho}_{ge;m,n-1} + i \Omega {\rho}_{ge;m,n-1}+ i g \sin\theta \Big[\sqrt{m} \rho_{ee; m-1,n-1} -\sqrt{n} \rho_{gg; m,n}\Big] \nonumber \\
&&- \left(\frac{1}{2} \Gamma_{\rm eff} + 2 \gamma_{\phi} \right) \, {\rho}_{ge;m,n-1},
\label{eq:r4}
\eea
and 
\bea
\dot{\rho}_{00;m,n} =&& (\bar{\Gamma}_{Lg}^s + \bar{\Gamma}_{Rg}^c ) \rho_{gg;m,n} + (\bar{\Gamma}_{Le}^c + \bar{\Gamma}_{Re}^s ) \rho_{ee;m,n} - (\Gamma_{Le}^c + {\Gamma}_{Lg}^s +{\Gamma}_{Rg}^c + \Gamma_{Re}^s) \rho_{00;m,n}.
\label{ground}
\eea
\end{widetext}
To close the equations, we also make use of the following two equations,
%
\bea
(\rho_{ph})_{m,n} &=& \rho_{00;m,n} + \rho_{gg;m,n} + \rho_{ee;m,n}, \\
(\rho_{ph})_{m-1,n-1}
&=& \rho_{00;m-1,n-1}+\rho_{gg;m-1, n-1}+\rho_{ee;m-1,n-1} \nonumber\\
\label{ph-norm}
\eea
We now employ the adiabatic approximation and solve for the steady state of the DQD. 
We first set $\dot{\rho}_{00;m,n}=0$ in Eq.~(\ref{ground}) and express $\rho_{gg;m,n}$ or $\rho_{ee;m,n}$ 
in terms of $(\rho_{ph})_{mn}$ using Eqs.~(\ref{ph-norm}).  
We then find, 
\bea
\rho_{00;m,n} &=& \frac{\bar{a}}{d+\bar{a}} (\rho_{ph})_{mn} + \frac{\bar{b}-\bar{a}}{d+\bar{a}} \rho_{ee;m,n} \nonumber \\
\rho_{00;m,n} &=& \frac{\bar{b}}{d+\bar{b}} (\rho_{ph})_{mn} + \frac{\bar{a}-\bar{b}}{d+\bar{b}} \rho_{gg;m,n}.
\eea
%
Here $\bar{a},\bar{b},d$ depend on the various parameters of the DQD, electronic and phononic baths;
their expressions are given in Appendix A. 
We next substitute  the solution for $\rho_{00;m,n}$ into  Eqs. (\ref{eq:r1})-(\ref{eq:r4})
to get closed set of equations, which can be written in the following matrix form,
\begin{widetext}
{\small
\be
\begin{pmatrix} \dot{\rho}_{gg;m,n} \\ \dot{\rho}_{ee;m-1,n-1} \\ \dot{\rho}_{eg;m-1,n} \\ \dot{\rho}_{ge;m,n-1}\end{pmatrix} =
  \begin{pmatrix} a_{11} &  0 & i g \sin \theta \sqrt{m} & -i g \sin \theta \sqrt{n} \\ 0 & a_{22} & -i g \sin \theta \sqrt{n} & i g \sin \theta \sqrt{m} \\  i g \sin \theta \sqrt{m} & -i g \sin \theta \sqrt{n} & a_{33} -i (\Omega - \omega_c) & 0 \\  -i g \sin \theta \sqrt{n} & i g \sin \theta \sqrt{m} & 0 & a_{33} + i (\Omega - \omega_c) \end{pmatrix}
 \begin{pmatrix} {\rho_{gg;m,n}} \\ {\rho_{ee;m-1,n-1}} \\ {\rho_{eg;m-1,n}} \\ {\rho_{ge;m,n-1}} \end{pmatrix} + 
\begin{pmatrix} b_1 p_m \\ b_2 (\rho_{ph})_{m-1,n-1}\\ 0 \\ 0\end{pmatrix}
\label{eq-matrix}
\ee
}
\end{widetext}
where once again $a_{11}$, $a_{22}$, $a_{33}$, $b_{1}$, $b_{2}$ are combination of parameters related to previously defined
constants, $\bar{a}$, $\bar{b}$, etc. Explicit expressions are given in the Appendix. 
In the steady state limit we set the left side of Eq.~(\ref{eq-matrix}) to zero and invert the matrix 
to obtain $\rho_{eg;m-1,m}$ and $\rho_{ge;m,m-1}$, as required by Eq.~(\ref{photon-pop}),
 which are solely expressed in terms of the cavity mode population, 
%
\bea
&&\rho_{eg;m-1,m} =
\nonumber \\
&&\frac{ i g \sin(\theta) \sqrt{m} \left(\frac{\tilde{A}}{2} p_{m-1} - \frac{\tilde{A}_b}{2} p_m \right) 
\left[ 1 + \frac{i (\Omega-\omega_c)}{a_{33}}\right]}{ 1 + \tilde C m g^2 \sin^2(\theta) + \frac{(\Omega-\omega_c)^2}{a_{33}^2}} 
\nonumber \\
&&\rho_{ge;m,m-1} = 
\nonumber \\
&&\frac{ i g \sin(\theta) \sqrt{m} \left(-\frac{\tilde{A}}{2} p_{m-1} + \frac{\tilde{A}_b}{2} p_m \right) 
\left[ 1 - \frac{i (\Omega-\omega_c)}{a_{33}}\right]}{ 1 + \tilde C m g^2 \sin^2(\theta) + \frac{(\Omega-\omega_c)^2}{a_{33}^2}}
\label{coherence}
\eea
%
The combined constants $\tilde A$,  $\tilde A_b$ and $\tilde C$ are given in the Appendix. 
These set of equations are substituted into Eq.~(\ref{photon-pop}) and we reach our central equation, 
%
\begin{eqnarray}
&&\frac{d}{dt} p_m =
 \frac{m \left[A p_{m-1} - A_b p_{m}\right]}{1 + m  C + \frac{(\Omega-\omega_c)^2}{a_{33}^2}} 
- \frac{(m+1) \left[A p_{m} - A_b p_{m+1}\right]}{1 + (m+1)  C+ \frac{(\Omega-\omega_c)^2}{a_{33}^2}} 
\nonumber \\ 
&&+  \kappa(1+ \bar{n}) \left[(m+1) p_{m+1} - m p_m\right] 
\nonumber \\
&&+\kappa \bar{n} \left[ m p_{m-1}- (m+1) p_m \right]. 
\label{central-eq}
\end{eqnarray}
The constants $A, A_b$ and $C$, 
expressed in terms of $a_{11}$, $a_{22}$, $a_{33}$, $b_{1}$, and $b_{2}$ are included in the Appendix. 
The small-letter constants combine the electronic decay rates and the phonon-induced energy relaxation constants. Note that,
$A$ and $A_b$ have the dimension of inverse time (rate) whereas $C$ is dimensionless. They quadratically depend on the light-matter coupling $g$.

\begin{figure}
\begin{center}
\includegraphics[scale=0.45]{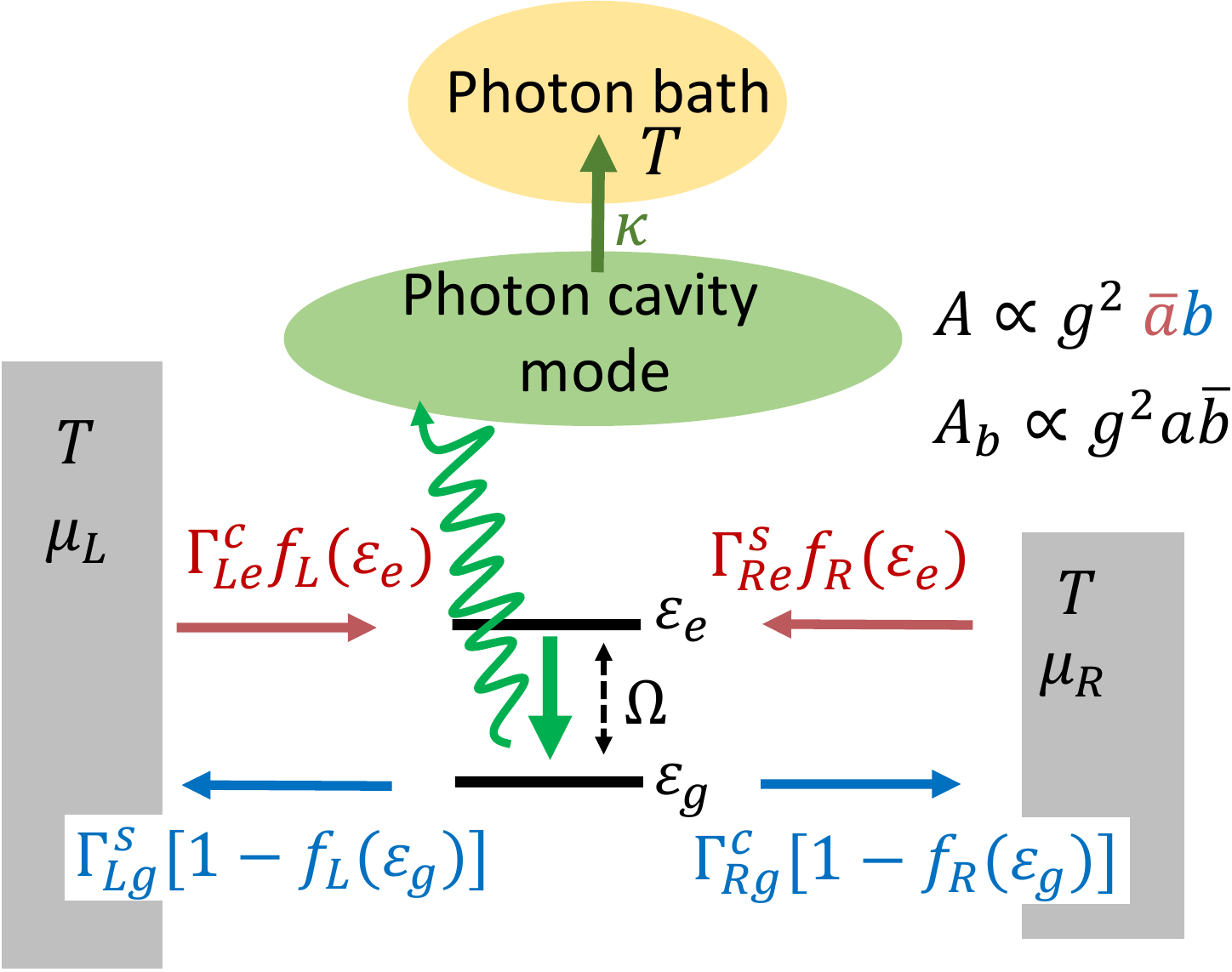}  
\caption{Schematic representation of charge transfer processes that contribute to photon gain, $A$.
The states of the DQD are depicted in the energy representation, $\epsilon_{g,e}$, with the gap $\Omega$
[see Eqs. (\ref{transformation}-\ref{eq:mattercavity})].
In the absence of phonons, $\bar a$ and $b$, which are defined in the Appendix, correspond to the rates of populating
the excited state (inward arrows) and rates of depleting the ground state (outward arrows), respectively.
The dissipation rate of cavity photons to the electronic medium is given by $A_b \propto \bar b a$;
these processes are not explicitly depicted here;
they can be visualized by reversing all the charge transfer processes (arrows) in the diagram.}
\label{scheme2}
\end{center}
\end{figure}

Equation (\ref{central-eq}), our central result, is quite compound, and
we now discuss the behavior of the photon statistics in steady state in two cases.
First we focus on the below-threshold regime and compare our result with the previously-obtained semiclassical prediction.
Then, we go back to the general result but simplify it at low temperature, thereby deriving closed expressions for photon statistics.
We further explain below through figure \ref{scheme2} the physical information contained in $A$ and $A_b$, which are
composite bias-driven light-matter rates determining the onset of masing in the system.


%

{\it Below threshold behavior.} 
For simplicity, we consider the resonance situation $\omega_c = \Omega$. 
When $C\ll 1$, which takes place when $g$ is smaller than bias-dependent 
electronic excitation and relaxation processes (see Appendix), 
one can ignore the denominator in the first two terms of Eq.~(\ref{central-eq}) and get
%
\begin{eqnarray}
\frac{d}{dt} p_m &=& m \big[A p_{m-1}- A_b p_{m}\big] 
\nonumber \\
&-& (m+1) \big[A p_{m} - A_b p_{m+1}\big]  
\nonumber \\ 
&+&  \kappa(1+ \bar{n}) \big[(m+1) p_{m+1} - m p_m\big] 
\nonumber \\ 
&+&\kappa \bar{n} \big[ m p_{m-1}- (m+1) p_m \big]. 
\end{eqnarray}
%
This equation can be solved in steady state by invoking the detailed balance principle \cite{slbook}, namely,
\begin{eqnarray}
m\big[A p_{m-1} -A_b p_m\big] + \kappa \big[{\bar n} m p_{m-1} - (1+\bar{n}) m p_m\big] =0. \nonumber \\
\end{eqnarray}
We then receive the solution for the photon mode population as
\be
p_m = \left[1- \frac{A + \kappa \bar{n}}{A_b + \kappa (1+ \bar{n})}\right] 
\left[ \frac{A+ \kappa \bar{n}}{A_b + \kappa (1+\bar{n})}\right]^m.
\label{sol-below}
\ee 
The normalization condition (or in other words, the validity of the solution) requires that 
\bea
&&\frac{A + \kappa \bar{n}}{A_b + \kappa ( 1+ \bar{n})} \leq 1 \nonumber \\ 
&\Rightarrow & A-A_b \leq \kappa.
\label{eq:cond}
\eea 
%
We conclude that Eq. (\ref{sol-below}) is valid as long as  $A-A_b \leq \kappa$, 
which precisely matches the threshold condition obtained using the semiclassical analysis 
in Eq.~(\ref{threshold-semi}). 
This is the so-called below-threshold regime.
The steady state photon mode distribution is given by an exponentially decaying incoherent thermal distribution,
$p_m \approx e ^{-\beta_{\rm eff} m}$,
with an effective temperature $T_{\rm eff}= 1/k_B \beta_{\rm eff}$ defined as 
\be T_{\rm eff} \equiv  \frac{\hbar \omega_c} { k_{B} \ln \left[\frac{A_b + \kappa (1+ \bar{n})}{A + \kappa \bar{n}}\right]}.
\ee 
As expected, $T_{\rm eff}$ reduces to the photon-bath temperature $T$ when the light-matter interaction is switched off,
with $A, A_b=0$. 

We can further generalize the threshold condition (\ref{eq:cond}) by relaxing the resonance requirement
$\omega_c=\Omega$. In the general case, the photon mode population is obtained by 
replacing $A \to \frac{A}{\left[1+ \frac{(\Omega-\omega_c)^2}{a_{33}^2}\right]}$ 
and $A_b \to \frac{A_b}{\left[1+ \frac{(\Omega-\omega_c)^2}{a_{33}^2}\right]}$ 
in Eq.~(\ref{sol-below}). This generalized threshold condition again 
matches with the corresponding semiclassical prediction. 

{\it General Solution.}
%
We go back to Eq. (\ref{central-eq}) and solve it,
\be
p_m = p_0 \prod_{j=1}^m \left[ \frac{\kappa \bar{n} + \frac{A}{1+ j C + \big(\frac{\Omega-\omega_c}{a_{33}}\big)^2}}{\kappa \big(1+\bar{n}\big) + \frac{A_b}{1+ j C + \big(\frac{\Omega-\omega_c}{a_{33}}\big)^2}} \right].
\ee
This solution describes both below- and above-threshold regimes.
Assuming a resonance setup, $\Omega=\omega_c$, and that the temperature is low
such that photon bath-induced excitation of the cavity are missing, $\bar{n} \approx 0$, 
one simply obtains 
\be
p_m = p_0 \prod_{j=1}^m \frac{A}{A_b + \kappa (1 + j C)}.
\label{gen-sol}
\ee
The maximum probability appears at  the number 
\be
m^* = \frac{1}{\kappa C} \big(A \!-\! A_b\!-\!\kappa \big).
\ee
The condition $m^*> 0$ corresponds to the above threshold regime; 
Again we find that $A-A_b \geq \kappa$ is the threshold condition. 

The population distribution in this lasing regime is sharply peaked about $m^*$, which is equivalent to the average
photon number $\langle m \rangle$, with the fluctuation about the average
given as $\sigma^2 = \frac{A}{\kappa C} = m^* + \frac{A_b + \kappa}{\kappa C}$.
The corresponding Fano factor is $\sigma^2/\langle m \rangle= 1 + \frac{A_b + \kappa}{A}$.
In the limit $A \gg A_b + \kappa$, the Fano factor reduces to 1,  which implies a Poisson distribution.

The physical content of $A$ and $A_b$ can be visualized via Fig.~\ref{scheme2},
where we sketch charge transfer processes that contribute to photon gain $A$. 
For simplicity, substrate phonons are ignored in this picture (compare Fig. \ref{scheme2} to  Fig. \ref{schematic}).
In the absence of phonons, $A\propto g^2 b\bar a$, 
with $\bar a$  being proportional to $\Gamma_{\alpha e}^c f_{\alpha}(E_e), \alpha = L,R$,
and $b$ related to the
removal of electrons from the ground state to both leads, $\Gamma_{\alpha g}^c [1- f_{\alpha}(E_g)]$,
see the Appendix for more details.
Altogether, $A$ corresponds to populating the electronic excited state $\epsilon_e$,
with the simultaneous depletion of electrons from the ground level $\epsilon_g$ to the metals. This electronic decay process
is complemented by the excitation of the cavity mode.

In contrast, $A_b\propto g^2 a\bar b$. This combination describes the reversed process to $A$:
It consists of electron transfer from the leads to the ground state, the simultaneous removal of electrons 
from the excited state to the leads, with the de-excitation of the cavity mode.
This term therefore corresponds to photon decay due to energy dissipation to the electron medium. 
Cavity photons further decay at a rate constant $\kappa$.
Altogether, the total dissipation rate is  $A_b + \kappa$, and masing is achieved only when photon 
gain overcomes the total dissipation, i.e. when we satisfy the condition  $A \geq A_b + \kappa$.
This situation takes place when the metals are voltage-biased far enough from equilibrium. 
Finally, we note that in the weak electron-phonon coupling limit considered in this work the presence of phonons does not 
fundamentally alter this picture.


\begin{center}
\begin{figure*}
\includegraphics[scale=0.57]{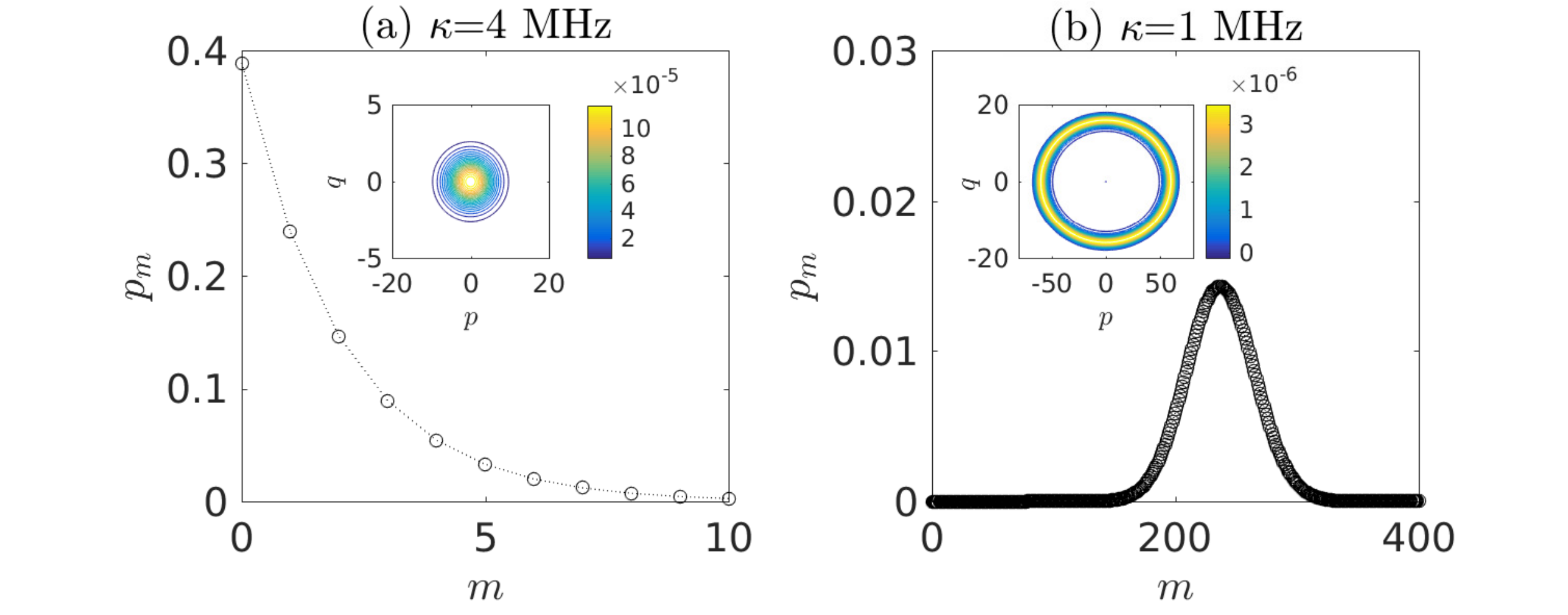}   
\includegraphics[scale=0.57]{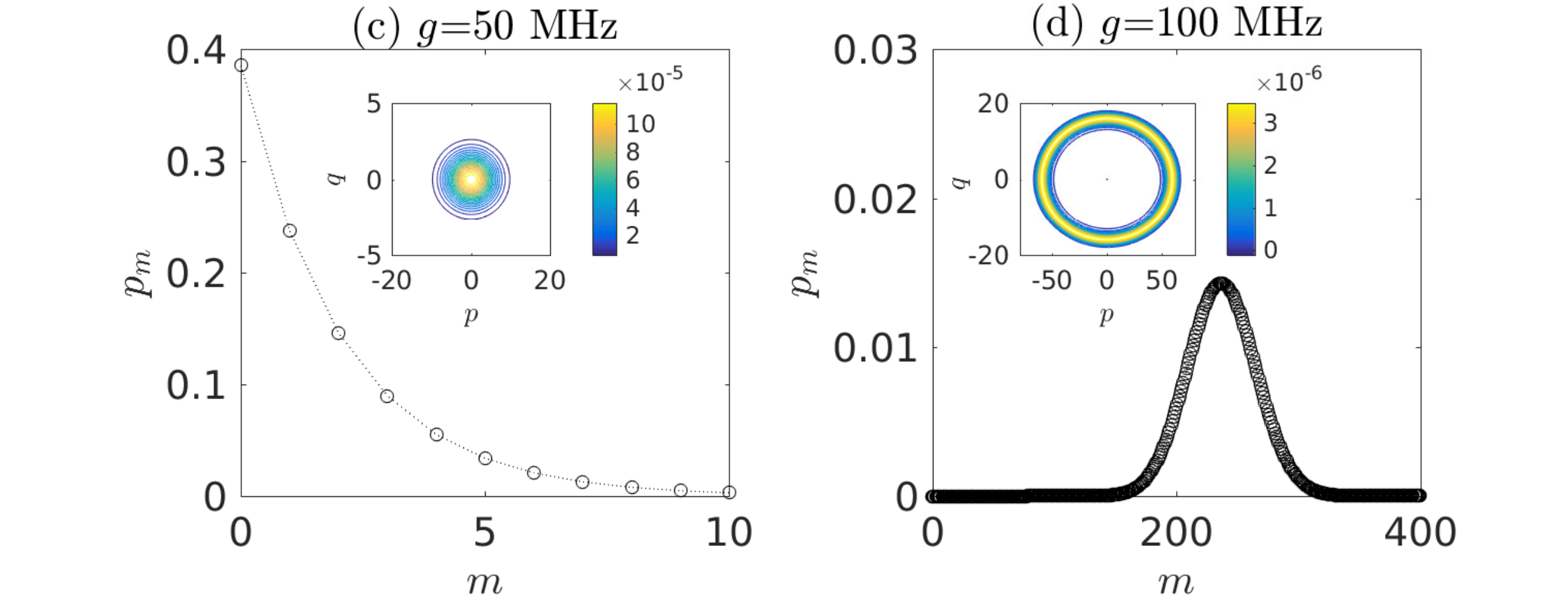}    
\includegraphics[scale=0.57]{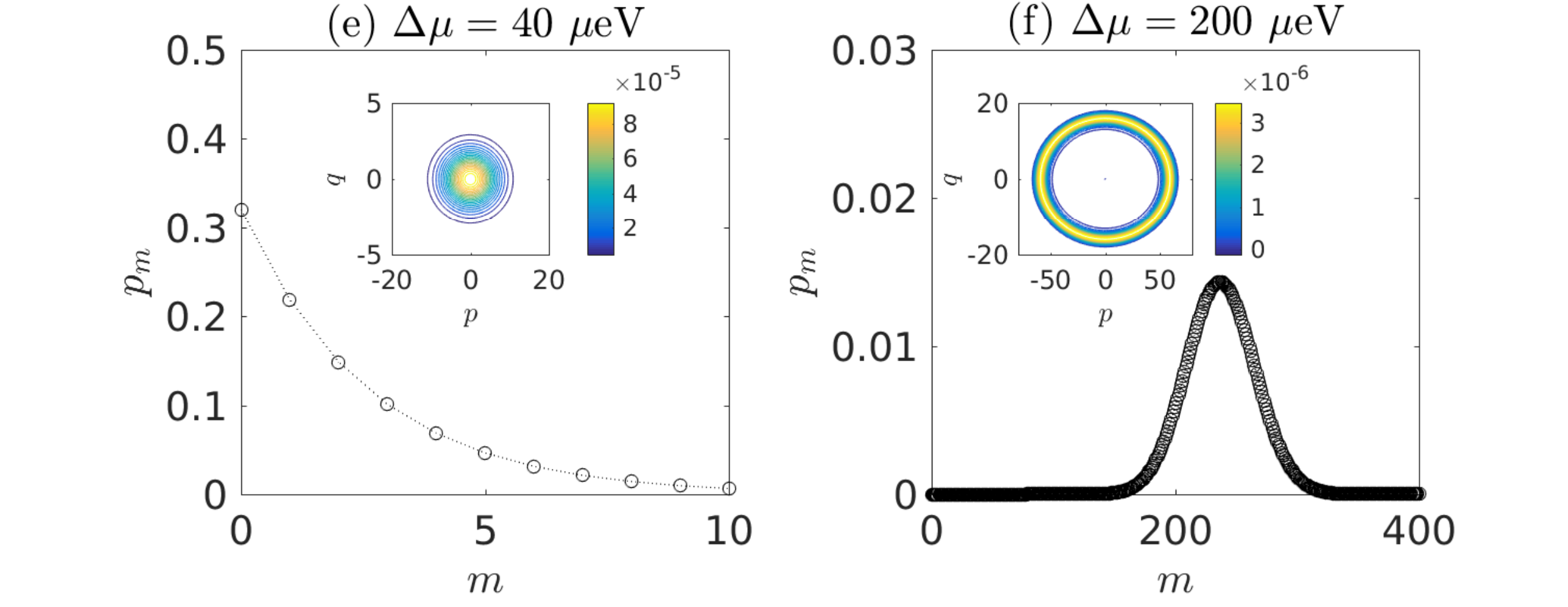}    
\includegraphics[scale=0.57]{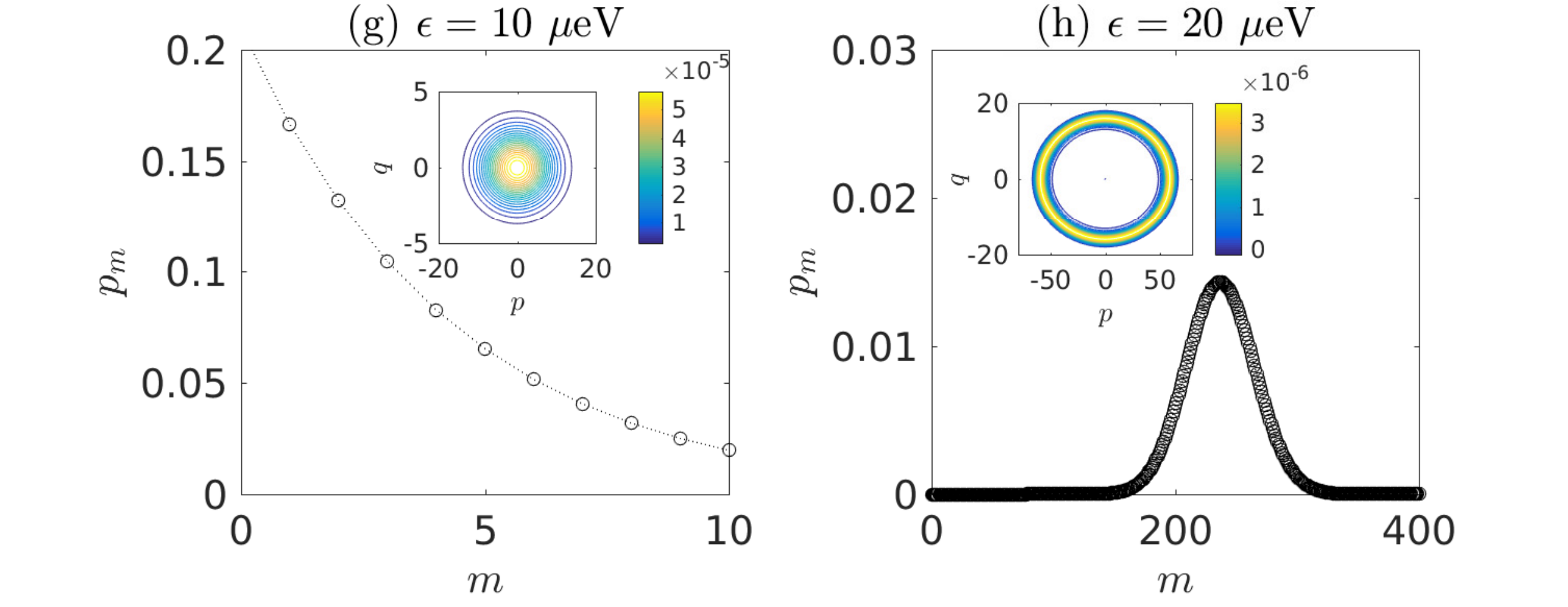}    
\caption{Photon statistics below (a,c,e,g) and above the masing threshold (b,d,f,h).
(a)-(b): $g=100$ MHz, $\Delta \mu =200$  $\mu${\rm eV}.
(c)-(d): $\kappa=1$ {\rm MHz}, $\Delta \mu$ =200 $\mu${\rm eV}.
(e)-(f): $g=100$ {\rm MHz} and $\kappa = 1$ {\rm MHz}.
(g)-(h): $g=100$ {\rm MHz} and $\kappa = 1$ {\rm MHz}, $\Delta \mu$=200 $\mu$eV.
Other parameters are  $t_c=16.4$ $\mu$eV, $\epsilon=20$ $\mu$eV, $\bar n=0$ and $\omega_c=\Omega$, see
also Table~1.
The inset in each panel shows the Wigner distribution in the $(q,p)$ plane.}
\label{photon_stat}
\end{figure*}
\end{center}


In Fig.~\ref{photon_stat}, we present the statistics of photons with various experimentally controllable parameters,
$\kappa$, $\Delta \mu$, $g$, $\epsilon$ based on Eq. (\ref{gen-sol}).
By tuning these parameters, we follow the crossover of the photon statistics from thermal  
(panels a, c, e, g) to Poissonian (panels b, d, f, h). 
We further plot the Wigner distribution in the $(q,p)$ plane for the corresponding density 
matrices,
\begin{eqnarray}
W(q,p) &=& \frac{1}{\pi \hbar} \int_{-\infty}^{\infty} dy e^{-\frac{i 2 y p}{\hbar}} \sum_{m=0}^{\infty} p_m \langle q-y|m\rangle \langle m | q +y \rangle \nonumber \\
&=&\frac{1}{\pi^{3/2}} \sum_{m=0}^{\infty} \!\frac{p_m}{2^m m!} \int_{-\infty}^{\infty}\! dy e^{-2i y p} \, e^{-(q^2 + y^2)} \nonumber \\&& \quad \quad  H_m(q-y) \, H_m(q+y),
\end{eqnarray}
with $p_m$  given by Eq.~(\ref{gen-sol}) and $H_m(x)$ the Hermite polynomials of order $m$. 
The Wigner function clearly exposes the two distinct regimes as far as photon statistics is concerned.
Given the control over the crossover between the different regimes, 
this setup can be potentially used as a quantum device 
that works both as a photon amplifier or a maser.

\section{Conclusions and Outlook}
\label{sec-summ}

In this paper, we analyzed the photonic properties of a QD-cQED system both below and above the masing threshold. 
Amplification and masing were achieved by driving the quantum dots out of equilibrium by voltage bias. 
We showed that the QD-cQED setup could serve as an excellent quantum device, both as a microwave amplifier and a maser 
as was demonstrated by recent experiments. 
We found a rich dependence of the threshold condition on experimentally tunable parameters,
such as the voltage bias, detuning, system-bath coupling. 
Our work also showed that these experimental knobs could be used to switch the device from a 
normal microwave amplifier to a maser. 

In both below and above the masing threshold we derived analytical expressions for the photon statistics 
and demonstrated via simulations the following distributions:
thermal (with an effective temperature) and Poissonian.
In addition, we found that the threshold condition obtained via the semiclassical approach precisely 
matched quantum calculations. 
It should be emphasized that this work critically extends our previous NEGF-based study \cite{longNEGF},  
which was limited to describing quantum-dot circuit-QED setups below the masing threshold.

Summarizing our results:
(i) We extended the Scully-Lamb quantum theory of the laser to the present setup, with a voltage bias driven electronic gain medium.
The coupling of the cavity to the DQD is included to all orders; photonic and phononic loss mechanisms were further evaluated.
(ii) We derived a threshold condition for masing using semiclassical and quantum theories. The two approaches yield
the same condition.
(iii) We investigated the transmission and statistics of emitted photons 
in relation to recent experiments and demonstrated the function of an amplifier
as well as the transition of the photon statistics from a thermal distribution into a Poissonian one.
Tunable parameters include bias voltage, DQD level splitting, DQD-cavity coupling, electron tunneling rates, 
dissipation rate and temperature. 



It remains a challenging task to understand the backaction of these photonic properties on the electronic degrees 
of freedom, and we aim to address this point in the future. 
Investigating lasing effects without working in the Born-Markov limit (thereby weak system-bath coupling) 
remains a significant challenge.  
Finally, scaling up the system to include multiple double-quantum dots will impact its masing performance,
a subject that we leave for future investigation. 

\begin{acknowledgments}
The authors thank Jason Petta and Takis Kontos for useful discussions. 
M. K. gratefully acknowledges the Ramanujan Fellowship
SB/S2/RJN-114/2016 from the Science and Engineering Research Board
(SERB), Department of Science and Technology, Government of India
and the hospitality of the Department of Chemistry at the  University of Toronto. 
B. K. A. gratefully acknowledges the start up support from IISER Pune. 
D.S. acknowledges the Natural Sciences and Engineering Research Council (NSERC) of Canada Discovery Grant
and the Canada Research Chairs Program.
\end{acknowledgments}

\renewcommand{\theequation}{A\arabic{equation}}
\setcounter{equation}{0}
\section*{Appendix: Definitions of constants used in Sec. \ref{photon-stat}}
We include here expressions for different constants defined in 
Sec.  \ref{photon-stat} on photon statistics, 
see  Eq.~(\ref{eq-matrix}) and Eq.~(\ref{coherence}).  We first define combinations of the 
electronic rate constants, as induced by the metal electrodes
\bea
\bar{a}&=& \bar{\Gamma}_{Lg}^s + \bar{\Gamma}_{Rg}^c, \quad \bar{b}= \bar{\Gamma}_{Le}^c + \bar{\Gamma}_{Re}^s, \nonumber \\
{a}&=& {\Gamma}_{Lg}^s + {\Gamma}_{Rg}^c, \quad {b}= {\Gamma}_{Le}^c + {\Gamma}_{Re}^s, \nonumber \\
d&=& \Gamma_{Le}^c + \Gamma_{Lg}^s + \Gamma_{Re}^s + \Gamma_{Rg}^c  = a + b  
\eea
We further define the following combination of phonon-bath and electron-bath induced processes,
\bea
a_{11} &=& (a -\gamma_{d}) \left(\frac{\bar{a}-\bar{b}}{\bar{b} + d}\right) - (\bar{a} + \gamma_{u} + \gamma_{d}), \nonumber \\
a_{22} &=& (b-\gamma_{u}) \left(\frac{\bar{b}-\bar{a}}{\bar{a} + d}\right) 
- (\bar{b} + \gamma_{u} + \gamma_{d}), \nonumber \\
a_{33} &=& -\left(\frac{1}{2} \Gamma_{\rm eff} + 2 \gamma_{\phi}\right),\nonumber \\
b_1 &=& \left[\gamma_{d} + \frac{(a -\gamma_{d}) \bar{b}}{\bar{b} + d} \right],\nonumber \\
b_2 &=& \left[\gamma_{u} + \frac{(b -\gamma_{u}) \bar{a}}{\bar{a} + d} \right].
\eea
The coefficients that appear in Eq.~(\ref{eq-matrix}) are 
\bea
\tilde{A} &=& \frac{ 2 b_2}{a_{22} a_{33}}, 
\quad \tilde{A}_b = \frac{ 2  b_1}{a_{11} a_{33}}, \quad \tilde{C}= \frac{2 (a_{11} + a_{22})}{a_{11} a_{22} a_{33}} \nonumber \\
A&=& g^2 \sin^2(\theta) \tilde{A}, 
\nonumber\\
A_b&=& g^2 \sin^2(\theta) \tilde{A}_b, 
\nonumber\\
\quad C&=& g^2 \sin^2(\theta) \tilde{C}. 
\eea
Note that these combinations depend in a non-trivial manner on the fundamental parameters of the c-QED setup. It is also important to notice that $A$ and $A_b$ have the dimension of inverse time and $C$ is dimensionless.

In the absence of phonons these expressions simplify to
\bea
\tilde A= \frac{2b\bar a}{b\bar a + \bar b \bar a + a \bar b}, \,\,\,\,
\tilde A_b= \frac{2a\bar b}{b\bar a + \bar b \bar a + a \bar b}.
\eea
Therefore,
\bea
\tilde A &\propto&  b\bar a 
=  \left(\Gamma_{Le}^c + \Gamma_{Re}^s \right) \left(  \bar{\Gamma}_{Lg}^s + \bar{\Gamma}_{Rg}^c\right)
\nonumber\\
 &=&\left[ \Gamma_{Le} f_L(\epsilon_e)\cos^2 \frac{\theta}{2} 
+ \Gamma_{Re} f_R(\epsilon_e)\sin^2 \frac{\theta}{2} \right] 
\nonumber\\
&\times&
\left[  \Gamma_{Lg}(1-f_L(\epsilon_g)) \sin^2 \frac{\theta}{2} 
+\Gamma_{Rg}(1-f_R(\epsilon_g)) \cos^2 \frac{\theta}{2}
\right]
\nonumber\\
\eea
and 
\bea
\tilde A_b&\propto& a \bar b = 
\left({\Gamma}_{Lg}^s + {\Gamma}_{Rg}^c\right)\left(  \bar{\Gamma}_{Le}^c + \bar{\Gamma}_{Re}^s \right)
\nonumber\\
 &=&\left[ \Gamma_{Lg} f_L(\epsilon_g)\sin^2 \frac{\theta}{2}
+ \Gamma_{Rg} f_R(\epsilon_g)\cos^2 \frac{\theta}{2} \right]
\nonumber\\
&\times&
\left[  \Gamma_{Le}(1-f_L(\epsilon_e)) \cos^2 \frac{\theta}{2}
+\Gamma_{Re}(1-f_R(\epsilon_e)) \sin^2 \frac{\theta}{2}
\right]
\nonumber\\
\eea 
Constructing $A$ and $A_b$ from $\tilde A$ and $\tilde A_b$, respectively,
we now recognize that $A$ embodies photon generation in the cavity, enabled by the bias-driven electronic system, while $A_b$
describes the decay of cavity photons by energy dissipation to the metals. 

We furthermore simplify $\tilde C$ in the absence of phonons,
\bea
\tilde C= \frac{\bar a+\bar b + 2a +2b}{(a\bar b + b \bar a+\bar a \bar b)(\bar a+\bar b)},
\eea
with $C\propto g^2\tilde C$. 
Therefore, the limit $C\ll 1$ corresponds to $g^2$ being much smaller than electronic 
processes in the metals, $g^2\ll \bar a b + \bar b a + \bar a \bar b$.


\begin{thebibliography}{99}








\bibitem{singleatom1}
Y.-Y. Liu, J. Stehlik, C. Eichler, X. Mi, T. R. Hartke, M. J. Gullans, J. M. Taylor, and J. R. Petta,  
Threshold dynamics of a semiconductor single atom maser,
Phys. Rev. Lett. \textbf{119}, 097702 (2017).


\bibitem{RevNori}
Z.-L. Xiang, S. Ashhab, J. Q. You, and F. Nori, 
Hybrid quantum circuits: Superconducting circuits interacting with other quantum systems, 
Rev. Mod. Phys. {\bf 85}, 623 (2013).


\bibitem{PNAS1}
G Kurizki, P. Bertet, Y. Kubo, K. Molmer, D. Petrosyan, P. Rabl, and J. Schmiedmayer, 
Proc. Natl. Acad. Sci. {\bf 112}, 3866 (2015).



\bibitem{kontoskondo}
M. R. Delbecq, V. Schmitt, F. D. Parmentier, N. Roch, J. J. Viennot, G. Fasve, B. Huard, C. Mora, A. Cottet, and T. Kontos,
Coupling a quantum dot, fermionic leads, and a microwave cavity on a chip,
Phys. Rev. Lett. {\bf 107}, 256804 (2011).

\bibitem{kontosnatcom}
M. R. Delbecq, L. E. Bruhat, J. J. Viennot, S. Datta, A. Cottet, and T. Kontos,
Photon-mediated interaction between distant quantum dot circuits,
Nat. comm. {\bf 4}, 1400 (2013).


\bibitem{kontosprx}
L. E. Bruhat, J. J. Viennot, M. C.  Dartiailh, 
M. M. Desjardins, T. Kontos, and A. Cottet,
Cavity photons as a probe for charge relaxation resistance and photon emission in a quantum dot coupled to 
normal and superconducting continua,
Phys. Rev. X {\bf 6}, 021014 (2016).

\bibitem{marco14}
M. Schiro and K. Le Hur, 
Tunable hybrid quantum electrodynamics from nonlinear electron transport,
Phys. Rev. B {\bf 89}, 195127 (2014).

\bibitem{kulqda}
J.H. Jiang, M. Kulkarni, D. Segal, and Y. Imry
Phonon-thermoelectric transistors and rectifiers,
Phys. Rev. B \textbf{92}, 045309 (2015).

\bibitem{kulqdb}
J. Lu, R. Wang, J.Ren, M. Kulkarni, J-H. Jiang,
Quantum-dot circuit-QED thermoelectric diodes and transistors,
Phys. Rev. B {\bf 99}, 035129  (2019).

\bibitem{KLH2015}
K. Le Hur, L. Henriet, A. Petrescu, K. Plekhnov, G. Roux, and M. Schiro, 
Many-body quantum electrodynamics networks: Non-equilibrium condensed matter physics with light, 
C. R. Physique {\bf 17}, 808 (2016).



\bibitem{pettamaserscience}
Y.-Y. Liu, J. Stehlik,  C. Eichler,  M. J. Gullans,  J. M. Taylor, and J. R. Petta, 
Semiconductor double quantum dot micromaser, 
Science {\bf 347}, 285 (2015).

\bibitem{pettamaserprl}
M. J. Gullans, Y. Y. Liu, J. Stehlik, J. R. Petta, and J. M. Taylor,
Phonon-assisted gain in a semiconductor double quantum dot maser,
Phys. Rev. Lett. {\bf 114}, 196802 (2015).

\bibitem{Petersson2012}
K. D. Petersson, L. W. McFaul, M. D. Schroer, M. Jung, J. M. Taylor, A. A. Houck, and J. R. Petta, 
Circuit quantum electrodynamics with a spin qubit,
Nature {\bf 490}, 380 (2012).

\bibitem{Frey2012}
T. Frey, P. J. Leek, M. Beck, A. Blais, T. Ihn, K. Ensslin, and A. Wallraff,
Dipole coupling of a double quantum dot to a microwave resonator,
Phys. Rev. Lett. {\bf 108}, 046807 (2012).


\bibitem{enslinprl}
A. Stockklauser, V. F. Maisi, J. Basset, K. Cujia, C. Reichl, W. Wegscheider, T. Ihn, A. Wallraff, and K. Ensslin,
Microwave emission from hybridized states in a semiconductor charge qubit,
Phys. Rev. Lett. {\bf 115}, 046802 (2015).


\bibitem{Toida2013}
H. Toida, T. Nakajima, and S. Komiyama,
Vacuum Rabi splitting in a semiconductor circuit QED system,
Phys. Rev. Lett. {\bf 110}, 066802 (2013).



\bibitem{Deng2013}
G.-W. Deng, D. Wei, J. R. Johansson, M.-L. Zhang, S.-X. Li, H.-O. Li, G. Cao, M. Xiao, T. Tu, G.-C. Guo, H.-W. Jiang, F. Nori, and G.-P. Guo,
Charge number dependence of the dephasing rates of a Graphene double quantum dot in a circuit QED architecture,
Phys. Rev. Lett. {\bf 115}, 126804 (2015).

\bibitem{gpg}
G.-W. Deng, L. Henriet, D. Wei, S.-X. Li, H.-O. Li, G. Cao, M. Xiao, G.-C. Guo, M. Schiro, K. Le Hur, and G.-P. Guo, 
A quantum electrodynamics Kondo circuit with orbital and spin entanglement,
arXiv:1509.06141. 


\bibitem{Viennot2013}
J. J. Viennot, M. R. Delbecq, M. C. Dartiailh, A. Cottet, and T. Kontos,
Out-of-equilibrium charge dynamics in a hybrid circuit quantum electrodynamics architecture,
Phys. Rev. B {\bf 89}, 165404 (2014).

\bibitem{spin-kontos}
J. J. Viennot, M. C. Dartiailh, A. Cottet, and T. Kontos, 
Coherent coupling of a single spin tomicrowave cavity photons,
Science \textbf{349}, 408 (2015).


\bibitem{tk2011}
A. Cottet, C. Mora, and T. Kontos, 
Mesoscopic admittance of a double quantum dot, 
Phys. Rev. B {\bf 83}, 121311 (2011).

\bibitem{petta2014}
Y.-Y. Liu, K. D. Petersson, J. Stehlik, J. M. Taylor, and J. R. Petta, 
Photon emission from a cavity-coupled double quantum dot,
Phys. Rev. Lett. {\bf 113}, 036801 (2014).

\bibitem{Kulkarni2014}
M. Kulkarni, O. Cotlet, and H. E. T\"ureci, 
Cavity-coupled double-quantum dot at finite bias: Analogy with lasers and beyond, 
Phys. Rev. B {\bf 90}, 125402 (2014).

\bibitem{Mora1} 
O. Dmytruk, M. Trif, C. Mora, and P. Simon,
Out-of-equilibrium quantum dot coupled to a microwave cavity,
Phys. Rev. B {\bf 93}, 075425 (2016).

\bibitem{Mora2} 
U. C. Mendes and C. Mora,
Electron-photon interaction in a quantum point contact coupled to a
microwave resonator,
Phys. Rev. B {\bf 93}, 235450 (2016).

\bibitem{QCL}
B. K Agarwalla, M. Kulkarni, S. Mukamel, and D. Segal,
Giant optical amplification in large-scale quantum dot circuit-QED systems,
Phys. Rev. B {\bf 94}, 121305(R) (2016).





\bibitem{scale1} 
D. M. Zajac, T. M. Hazard, X. Mi, E. Nielsen, and J. R. Petta, 
Scalable gate architecture for a one-dimensional array of semiconductor spin qubits,
Phys. Rev. Appl. {\bf 6}, 054013 (2016).

\bibitem{scale2} 
A. R. Mills, D. M. Zajac, M. J. Gullans, F. J. Schupp, T. M. Hazard, and J. R. Petta, 
Shuttling a single charge across a one-dimensional array of silicon quantum dots,
arXiv: 1809.03976. 

\bibitem{siv1}
G. Burkard and J. R. Petta, 
Non-local transport properties of nanoscale conductor-microwave cavity systems,
Phys. Rev. B \textbf{94}, 195305 (2016).

\bibitem{pd1}
V. Gudmundsson, N. R. Abdulla, A. Sitek, H. Goan, C. Tang, A. Manolescu, 
Electroluminescence Caused by the Transport of Interacting Electrons through Parallel Quantum Dots in a Photon Cavity,  
Ann. Phys., {\bf 530}, 1700334 (2018). 

\bibitem{pd2}
V. Gudmundsson, N. R. Abdulla, A. Sitek, H. Goan, C. Tang, A. Manolescu, 
Current correlations for the transport of interacting electrons through parallel quantum dots in a photon cavity, 
Phys. Lett. A, {\bf 382}, 1672 (2018).

\bibitem{phaselock1}
Y.-Y. Liu, T. R. Hartke, J. Stehlik, and J. R. Petta,
Phase locking of a semiconductor double-quantum-dot single-atom maser,
Phys. Rev. A {\bf 96}, 053816 (2017).



\bibitem{Ryd} D. Meschede, H. Walther, and G. M\"uller, 
One-atom maser,
Phys. Rev. Lett. {\bf 54}, 551 (1985).



\bibitem{art1} J. McKeever, A. Boca, A. D. Boozer, J. R. Buck, and H. J. Kimble, 
Experimental realization of a one-atom laser in the regime of strong coupling, 
Nature (London) {\bf 425}, 268 (2003).

\bibitem{art2} 
S. Reitzenstein, T. Heindel, C. Kistner, A. Rahimi-Iman, C. Schneider, S. H\"ofling, and A. Forchel, 
Low threshold electrically pumped quantum dot-micropillar lasers, 
Appl. Phys. Lett. {\bf 93}, 061104 (2008).

\bibitem{art3} 
F. Dubin, C. Russo, H. G. Barros, A. Stute, C. Becher, P. O. Schmidt, and R. Blatt, 
Quantum to classical transition in a single-ion laser, 
Nat. Phys. {\bf 6}, 350 (2010).

\bibitem{art4} S. Haroche, 
Nobel lecture: Controlling photons in a box and exploring the quantum to classical boundary, 
Rev. Mod. Phys. {\bf 85}, 1083 (2013).



\bibitem{super1} 
O. V. Astafiev, A. A. Abdumalikov, A. M. Zagoskin, Y. A. Pashkin, Y. Nakamura, and J. S. Tsai, 
Ultimate on-chip quantum amplifier, 
Phys. Rev. Lett. {\bf 104}, 183603 (2010).

\bibitem{super2} O. Astafiev, K. Inomata, A. O. Niskanen, T. Yamamoto, Y. A. Pashkin, Y. Nakamura, and J. S. Tsai, 
Single artificial-atom lasing,
Nature (London) {\bf 449}, 588 (2007).



\bibitem{longNEGF}
B. K. Agarwalla, M. Kulkarni, S. Mukamel, and D. Segal, 
Tunable photonic cavity coupled to a voltage-biased double quantum dot system: Diagrammatic NEGF approach, 
Phys. Rev. B {\bf 94}, 035434 (2016).


\bibitem{slbook}
M. O. Scully and M. Suhail Zubairy,  Quantum Optics, Cambridge University Press.

\bibitem{Agarwal}
S.-W. Li, M. B. Kim, G. S. Agarwal, and M. O. Scully,
Quantum statistics of a single-atom Scovil-Schulz-DuBois heat engine,
Phys. Rev. A {\bf 96}, 063806 (2017).


\bibitem{Vavilov-fcs}
C. Xu and M. G. Vavilov, 
Full counting statistics of photons emitted by a double quantum dot,
Phys. Rev. B \textbf{88}, 195307 (2013).


\bibitem{Nori-bistable} 
N. Lambert, F. Nori, and C. Flindt, 
Bistable photon emission from a solid-state single-atom laser,
Phys. Rev. Lett {\bf 115}, 216803 (2015).


\bibitem{ss2}
C. Bergenfeldt and P. Samuelsson,
Microwave quantum optics and electron transport through a metallic dot strongly coupled to a transmission line cavity,
Phys. Rev. B \textbf{85}, 045446 (2012).


\bibitem{ss1}
C. Bergenfeldt and P. Samuelsson,
Non-local transport properties of nanoscale conductor-microwave cavity systems,
Phys. Rev. B \textbf{87}, 195427 (2013).

\bibitem{Schon_noise} 
J. Jin, M. Marthaler,  P.Q. Jin, D. Golubev, and G. Sch\"on, 
Noise spectrum of a quantum dot-resonator lasing circuit,
New J. Phys {\bf 15}, 025044 (2013).


\bibitem{siv2}
X. Mi, Csaba G. Peterfalvi, G. Burkard, and J. R. Petta,
High-resolution valley spectroscopy of Si quantum dots,
Phys. Rev. Lett. \textbf{119}, 176803 (2017).

\bibitem{Schon}
C. Karlewski, A. Heimes, and G. Sch\"on, 
Lasing and transport in a multilevel double quantum dot system coupled to a microwave oscillator, 
Phys. Rev. B {\bf 93} 045314 (2016).

\bibitem{Child} A. Childress, S. Sørensen, and M. D. Lukin, 
Mesoscopic cavity quantum electrodynamics with quantum dots, 
Phys. Rev. A {\bf 69}, 042302 (2004).

\bibitem{Jin} P.-Q. Jin, M. Marthaler, J. H. Cole, A. Shnirman, and G. Sch\"on, 
Lasing and transport in a quantum-dot resonator circuit, 
Phys. Rev. B {\bf 84}, 035322 (2011).

\bibitem{Muller} 
C. M\"uller and T. M. Stace, 
Deriving Lindblad master equations with Keldysh diagrams: Correlated gain and loss in higher,
Phys. Rev. A {\bf 95}, 013847 (2017). 




\bibitem{tb1}
L. D. Contreras-Pulido, C. Emary, T. Brandes, and R. Aguado, 
Non-equilibrium correlations and entanglement in a semiconductor hybrid circuit-QED system,
New J. Phys. \textbf{15}, 095008 (2013).


\bibitem{petta-phonon}
T. R. Hartke, Y.-Y. Liu, M. J. Gullans, and J. R. Petta, 
Microwave detection of electron-phonon onteractions in a cavity-coupled double quantum dot,
Phys. Rev. Lett {\bf 120}, 097701 (2018).

\bibitem{weber-phonon}
C. Weber, A. Fuhrer, C. Fasth, G. Lindwall, L. Samuelson, and A. Wacker, 
Probing confined phonon modes by transport through a nanowire double quantum dot,
Phys. Rev. Lett. {\bf 104}, 036801 (2010).

\end{thebibliography}
\end{document}